\documentclass[universe,review,accept,moreauthors,pdftex,12pt,a4paper]{mdpi} 
\lastpage{x}
\doinum{10.3390/------}
\pubvolume{xx}
\pubyear{2015}
\history{Received: xx / Accepted: xx / Published: xx}

\usepackage{amsmath}
\usepackage{amssymb}

\newcommand{\lp}{\left(}
\newcommand{\rp}{\right)}
\newcommand{\lb}{\left[}
\newcommand{\rb}{\right]}

\def\bea{\begin{eqnarray}}
\def\eea{\end{eqnarray}}

\newcommand{\ba}{\begin{eqnarray}}
\newcommand{\ea}{\end{eqnarray}}
\newcommand{\be}{\begin{equation}}
\newcommand{\ee}{\end{equation}}

\newcommand{\al}{\alpha}
\newcommand{\bt}{\beta}

\newcommand{\ka}{\kappa}

\newcommand{\R}{\mathcal{R}}

\newcommand{\OA}{\Omega_A}
\newcommand{\e}{\rm e}

\newcommand{\F}{\OA+\phi}
\newcommand{\Fp}{\lp\OA+\phi\rp}
\newcommand{\f}{f^{(0)}_{,R\R}}
\newcommand{\Q}{f^{(0)}_{,\hat Q}}
\newcommand{\Frr}{f^{(0)}_{,RR}}

\Title{Hybrid metric-Palatini gravity}

\Author{Salvatore Capozziello $^{1,2,3}$, Tiberiu Harko $^4$, Tomi S. Koivisto $^{5}$, Francisco S.N.~Lobo $^{6,}$*, and Gonzalo J. Olmo $^{7}$}

\address{%
$^{1}$ Dipartimento di Fisica, Universit\`{a} di Napoli ``Federico I'', Napoli, Italy\\
$^{2}$ INFN Sez. di Napoli, Compl. Univ. di Monte S. Angelo, Edificio G, Via Cinthia, I-80126, Napoli, Italy, E-Mail: capozzie@na.infn.it \\
$^{3}$ Gran Sasso Science Institute (INFN), Viale F. Crispi /, I-67100, L'Aquila, Italy.\\
$^{4}$ Department of Mathematics, University College London, Gower Street, London WC1E 6BT, United Kingdom, E-Mail: t.harko@ucl.ac.uk\\
$^{5}$ Nordita, KTH Royal Institute of Technology and Stockholm 
University, Roslagstullsbacken 23, SE-10691 Stockholm, Sweden, E-Mail: tomi.koivisto@nordita.org\\
$^{6}$ Instituto de Astrof\'{\i}sica e Ci\^{e}ncias do Espa\c{c}o, Universidade de Lisboa, Edif\'{\i}cio C8, Campo Grande, PT1749-016 Lisbon, Portugal, E-Mail: fslobo@fc.ul.pt\\
$^7$ Departamento de F\'{i}sica Te\'{o}rica and IFIC, Centro Mixto Universidad de
Valencia - CSIC. Universidad de Valencia, Burjassot-46100, Valencia, Spain, E-Mail: gonzalo.olmo@csic.es}

\corres{E-Mail: fslobo@fc.ul.pt; Tel: +351 217 500 055; Fax:+351 217 500 977.}

\abstract{
Recently, the phenomenology of  $f(R)$ gravity has been scrutinized motivated by the possibility  to account for the self-accelerated cosmic expansion without invoking dark energy sources. Besides,  this kind of modified gravity is capable of addressing the dynamics of several self-gravitating systems alternatively to the presence of dark matter.  It has been established that both metric and Palatini versions of these theories have interesting features but also manifest severe and different downsides. A hybrid combination of theories, containing elements from both these two formalisms, turns out to be also very successful accounting for the observed phenomenology and is able to avoid some drawbacks of the original approaches. This article reviews the formulation of this hybrid metric-Palatini approach and its main achievements in passing the local tests and in applications to astrophysical and cosmological scenarios, where it provides a unified approach to the problems of dark energy and dark matter.}


\begin{document}


\section{Introduction}

One hundred years ago, Albert Einstein completed the mathematical formulation of his revolutionary view of the gravitational interactions in terms of curved space-time. The spirit, elegance, and experimental successes of the original theory has captivated the international scientific community and the theory has been accepted as the {\it standard model} for gravity \cite{bah}. At its centennial, pushed by new observational evidences, the theory is at a dramatic crossroad. Its continuation as the reference gravitational framework will imply that the universe is mainly composed by exotic sources of matter and energy whose existence is purely inferred from their gravitational effects at the largest astrophysical and cosmological scales  \cite{expansion1,expansion2, expansion3, expansion4, expansion5, expansion6,expansion7}. However, if such sources are not detected in any direct way, then we might be facing a failure of one of the most original theories of the twentieth century. The important implications of the two opposed alternatives, i.e., the search for unknown dark side constituents or the revision of gravitational theory,  demand a careful scrutiny of the different possible scenarios. In this work, we consider the second case, namely, the situation in which the gravitational dynamics may depart from that predicted by Einstein's theory of General Relativity (GR) at ultraviolet and infrared scales. We mainly focus on the latter regime.

Given the success of GR at relatively short scales (such as the Solar System, stellar models, or compact binary systems), the idea that modified dynamics could arise at larger scales has been investigated in much detail over the last years. Theories in which the gravitational action consists
of more general combinations of curvature invariants than the pure
Einstein-Hilbert term have been investigated with special emphasis
\cite{fRgravity1,fRgravity2, fRgravity3, fRgravity4, fRgravity5, fRgravity6, fRgravity7, fRgravity8, fRgravity9}. From these investigations it was soon noticed that the usual metric formulation of alternative theories of gravity is generically different from its Palatini (or metric-affine) counterpart (see \cite{Olmo:2011uz} for a recent
review on the Palatini approach). Whereas the metric approach typically leads to higher-order derivative equations, in the Palatini formulation the resulting field equations are always second-order. The appealing character of the second-order equations of the Palatini formalism, however, is accompanied by certain algebraic relations between the matter fields and the affine connection, which is now determined by a set of equations coupled to the matter fields and the metric. The case of $f(R)$ theories is particularly useful to illustrate the differences between these two approaches. In the metric formulation, the object $\phi\equiv {\rm d}f/{\rm d}R$ behaves as a dynamical scalar field, which satisfies a second-order equation with self-interactions that depend on the form of the Lagrangian $f(R)$. In order to have an impact at large astrophysical and cosmological scales, the scalar field $\phi$ should have a very low mass, implying a long interaction range. It is well known, however, that light scalars do have an impact at shorter scales, where their presence is strongly constrained by laboratory and Solar System observations unless some kind of screening mechanism is invoked \cite{screen1,screen2,screen3,screen4,screen5}. In the Palatini case, a scalar-tensor representation is also possible, but with the scalar field satisfying an algebraic rather than a differential equation. One then finds that the scalar $\phi$ turns out to be an algebraic function of the trace of the stress-energy tensor of the matter, $\phi=\phi(T)$, which may lead, in models of late-time cosmic speed-up, to undesired gradient instabilities at various contexts, as has been shown by studies of cosmological perturbations \cite{Koivisto:2005yc,Koivisto:2006ie} and atomic physics \cite{Olmo:2008ye, Olmo:2006zu}.

In this article we will review the {\it hybrid} variation of these theories, in which the (purely metric) Einstein-Hilbert action is supplemented with (metric-affine) correction terms constructed \`a la Palatini \cite{Harko:2011nh,IJMPD}. 
Given that metric and Palatini $f(R)$ theories allow the construction of simple extensions of GR with interesting properties and, at the same time, suffer from different types of drawbacks, we initiated a program to establish bridges between these two seemingly disparate approaches hoping to find ways to cure or improve their individual deficiencies. For that purpose, in a number of works we considered a hybrid combination of metric and Palatini elements to construct the gravity Lagrangian and found that viable models sharing properties of both formalisms are possible. An interesting aspect of these theories is the possibility to generate long-range forces without entering into conflict with local tests of gravity and without invoking any kind of screening mechanism (which would however require that at the present time the cosmological evolution reduces to GR.). The possibility of expressing these hybrid $f(R)$ metric-Palatini theories using a scalar-tensor representation simplifies the analysis of the field equations and the construction of solutions. 
In some sense, considering a theory like $R+f({\cal R})$ means that we retain all the positive results of GR, represented by the Einstein-Hilbert part of the action $R$, while the further ``gravitational budget'' is endowed in the metric-affine $f({\cal R})$ component.  In fact it is well known that metric-affine and purely metric formalisms coincide in GR, i.e., considering the action $R$.  On the  contrary, the two formalisms lead to different results considering more generic functions $f({\cal R})$ \cite{Olmo:2011uz}.

A related approach to study $f(R)$ theories that interpolate between the metric and Palatini families is that of the so-called C-theories proposed in \cite{Amendola:2010bk,Koivisto:2013xza}. 
There the spacetime connection is associated to the metric $\hat g_{\mu\nu}=C(\R)g_{\mu\nu}$ that is conformally related to the spacetime metric $g_{\mu\nu}$, but the relation may depend upon the a scalar curvature $\R$. This framework contains the metric, $C(\R)=1$, and the Palatini, $C(\R)=f'(R)$, formalisms as special limits, and one also finds that even when $f(\R)=\R$ physically distinct theories are possible. For further studies on variations of variational principles see e.g. \cite{vari0,varia,vari1,varib,vari2,vari3,vari4}.

Other extensions of the $f(R)$ framework modify the coupling of matter to gravity 
by defining an action which depends linearly \cite{Koivisto:2005yk} or nonlinearly upon the matter
Lagrangian \cite{Allemandi:2005qs,Olmo:2014sra,Harko:2010mv,Bertolami:2007gv,Bertolami:2007vu,Bertolami:2008ab,BPHL,HKL}, or its trace
\cite{Harko:2011kv, Haghani:2013oma, Odintsov:2013iba}. The new couplings generally induce non-geodesic motion mediated by an extra force orthogonal to the four-velocity \cite{Harko:2012ve}, which may have nontrivial effects  already in flat Minkowski space. Instabilities due to new nonlinear interactions within the matter sector are thus common in these theories \cite{stabi1,stabi2}. We note, from this perspective, that in the hybrid metric-Palatini approach, considered in this article, it can be expected that such instabilities in the matter sector are absent, because the usual conservation laws are satisfied.

In this paper, we review the formulation and the main applications of  hybrid gravity models in late-time cosmological  and astrophysical scenarios. The article is organized in three main parts considering the general formalism, the cosmology and the astrophysical applications.  
In Section 2, we start the discussion considering the action and the field equations of the hybrid metric-Palatini formalism. In particular, we discuss the scalar-tensor representation, the Cauchy problem, and more general hybrid theories than $R+f({\cal R})$. Section 3 is devoted to hybrid-gravity cosmology. We derive the Friedmann equations, construct the related dynamical system, and briefly consider some solutions. Furthermore, we analyse the cosmological perturbations in order to understand structure formation in these theories. We focus on the evolution of perturbations in the matter dominated era and vacuum fluctuations relevant to inflation. The weak field behaviour that is crucial for the Solar system precision tests of gravity is considered in Section 4, where we also discuss the galactic phenomenology and the astrophysical applications of hybrid gravity as an alternative to dark matter. In particular, we study the stellar dynamics and the theory of orbits, the generalization of virial theorem, the flat rotation curves of spiral galaxies, and the galactic clusters starting from the relativistic Boltzmann equation for collisionless systems of particles. 
The conclusions are drawn in Section 5.

\section{Hybrid metric-Palatini gravity: The general formalism}

Let us start our considerations by giving the basic features of the theory. In particular, we discuss the action and the field equations both in the so called $f(X)$ and its equivalent scalar-tensor representations in both the Jordan and the Einstein frames. Next, the well-formulation and well-posedness of the Cauchy problem is considered. Finally, arbitrary hybrid gravity theories constructed from the metric and an independent connection are explored and it is shown that the special $f(X)$ models avoid otherwise generic pathologies. 

\subsection{Action and gravitational field equations}

The action is specified as \cite{Harko:2011nh,Capozziello:2012ny}
\begin{equation} \label{eq:S_hybrid}
S= \frac{1}{2\kappa^2}\int {\rm d}^4 x \sqrt{-g} \left[ R + f(\R)\right] +S_m \ ,
\end{equation}
where $S_m$ is the matter action, $\kappa^2\equiv 8\pi G$, $R$ is
the Einstein-Hilbert term, $\R  \equiv g^{\mu\nu}\R_{\mu\nu} $ is
the Palatini curvature, defined in terms of
an independent connection $\hat{\Gamma}^\alpha_{\mu\nu}$  as
\be \R
\equiv  g^{\mu\nu}\R_{\mu\nu} \equiv g^{\mu\nu}\lp
\hat{\Gamma}^\alpha_{\mu\nu , \alpha}
       - \hat{\Gamma}^\alpha_{\mu\alpha , \nu} +
\hat{\Gamma}^\alpha_{\alpha\lambda}\hat{\Gamma}^\lambda_{\mu\nu} -
\hat{\Gamma}^\alpha_{\mu\lambda}\hat{\Gamma}^\lambda_{\alpha\nu}\rp\,,\label{r_def}
\ee 
that generates the Ricci curvature tensor $\R_{\mu\nu}$ as
\begin{equation}
\R_{\mu\nu} \equiv \hat{\Gamma}^\alpha_{\mu\nu ,\alpha} -
\hat{\Gamma}^\alpha_{\mu\alpha , \nu} +
\hat{\Gamma}^\alpha_{\alpha\lambda}\hat{\Gamma}^\lambda_{\mu\nu}
-\hat{\Gamma}^\alpha_{\mu\lambda}\hat{\Gamma}^\lambda_{\alpha\nu}\,.
\end{equation}

Varying the action given by Eq. (\ref{eq:S_hybrid}) with respect to the metric, one obtains the following gravitational field equation  
\be
\label{efe} G_{\mu\nu} +
F(\R)\R_{\mu\nu}-\frac{1}{2}f(\R)g_{\mu\nu} = \ka^2 T_{\mu\nu}\,,
\ee
where the matter stress-energy tensor is defined as usual,
 \be \label{memt}
 T_{\mu\nu} \equiv -\frac{2}{\sqrt{-g}} \frac{\delta
 (\sqrt{-g}\mathcal{L}_m)}{\delta(g^{\mu\nu})}.
 \ee
Varying the action with respect to the independent connection $\hat{\Gamma}^\alpha_{\mu\nu}$, one then finds as the solution to the resulting equation of motion that $\hat{\Gamma}^\alpha_{\mu\nu}$ is compatible with the metric
$F(\R)g_{\mu\nu}$, conformally related to the physical metric $g_{\mu\nu}$, with the conformal factor given by $F(\R) \equiv df(\R)/d\R$. This implies that
\ba
\label{ricci} \R_{\mu\nu} & = & R_{\mu\nu} +
\frac{3}{2}\frac{1}{F^2(\R)}F(\R)_{,\mu}F(\R)_{,\nu}
  - \frac{1}{F(\R)}\nabla_\mu F(\R)_{,\nu} -
\frac{1}{2}\frac{1}{F(\R)}g_{\mu\nu}\nabla_\alpha \nabla^\alpha F(\R)\,. \ea 
The Palatini curvature, $\R$, can be
obtained from the trace of the field equation (\ref{efe}), which
yields \be \label{trace} F(\R)\R -2f(\R)= \ka^2 T +  R \equiv X\,.
\ee
Note that we can express $\R$ algebraically in terms of $X$ if the form of $f(\R)$ allows analytic solutions. The variable $X$ measures how much  the theory deviates from the general relativity trace equation $R=-\ka^2 T$. These two observations shed light upon the structure of the theory and the central part that the ''trace deviation'' $X$ plays in it. It is for this reason the $R+f(\R)$ hybrid metric-Palatini theories have also been called simply the ''$f(X)$ theories''.

Indeed, we can express the field equation (\ref{efe}) solely in terms of the metric and $X$ as
\ba \label{efex} 
G_{\mu\nu} & = &
\frac{1}{2}f(X)g_{\mu\nu}- F(X)R_{\mu\nu}  +  F'(X)
\nabla_{\mu}X_{,\nu}
     + \frac{1}{2}\lb F'(X)\nabla_\alpha \nabla^\alpha X + F''(X)\lp \partial X\rp^2 \rb
g_{\mu\nu}
    \nonumber  \\
&&+ \lb F''(X)-\frac{3}{2}\frac{\lp
F'(X)\rp^2}{F(X)}\rb X_{,\mu}X_{,\nu} + \ka^2 T_{\mu\nu}\,. \ea
Note that $(\partial X)^2=X_{,\mu}X^{,\mu}$. The trace of the field equations is now \ba \label{trace2} F'(X)\nabla_\alpha \nabla^\alpha X + \lb
F''(X)-\frac{1}{2}\frac{\lp F'(X)\rp^2}{F(X)}\rb \lp \partial
X\rp^2+
 \frac{1}{3}\left[ X + 2f(X)-F(X)R\right]= 0 \,,
\ea  while the relation between the metric  scalar curvature  $R$ and
the Palatini scalar curvature  $\R$ is   
\be \label{ricciscalar} \R(X) =
R+\frac{3}{2}\lb \lp\frac{F'(X)}{F(X)}\rp^2-2\frac{\nabla_\alpha \nabla^\alpha
F(X)}{F(X)}\rb\,, \ee 
which can be obtained by contracting Eq.~(\ref{ricci}). 

\subsection{Scalar-tensor representation}
\label{sts}

Like in the pure metric and Palatini cases \cite{Olmo:2005zr,Olmo:2005hc}, the action (\ref{eq:S_hybrid}) for the hybrid metric-Palatini theory can be turned into that of a
scalar-tensor theory by introducing an auxiliary field $A$ such
that
\begin{equation} \label{eq:S_scalar0}
S= \frac{1}{2\kappa^2}\int {\rm d}^4 x \sqrt{-g} \left[\OA R + f(A)+f_A(\R-A)\right] +S_m \ ,
\end{equation}
where $f_A\equiv df/dA$ and we have included a coupling constant $\OA$ for generality. Note that $\OA=1$ for the original hybrid metric-Palatini theory \cite{Harko:2011nh}. Rearranging the terms and defining $\phi\equiv f_A$, $V(\phi)=A f_A-f(A)$,
Eq. (\ref{eq:S_scalar0}) becomes
\begin{equation} \label{eq:S_scalar1}
S= \frac{1}{2\kappa^2}\int {\rm d}^4 x \sqrt{-g} \left[\OA R + \phi\R-V(\phi)\right] +S_m \ .
\end{equation}
It is easy to see that this action is equivalent to our original starting point (\ref{eq:S_hybrid}).
Variation of the above action with respect to the metric, the scalar $\phi$ and the connection leads to the field equations
\begin{eqnarray}
\OA R_{\mu\nu}+\phi \R_{\mu\nu}-\frac{1}{2}\left(\OA R+\phi\R-V\right)g_{\mu\nu}&=&\kappa^2 T_{\mu\nu}\,,
\label{eq:var-gab}\\
\R-V_\phi&=&0 \label{eq:var-phi} \,, \\
\hat{\nabla}_\alpha\left(\sqrt{-g}\phi g^{\mu\nu}\right)&=&0 \,, \label{eq:connection}\
\end{eqnarray}
respectively. 

The solution of Eq.~(\ref{eq:connection}) implies that the independent connection is the Levi-Civita connection of
a metric $h_{\mu\nu}=\phi g_{\mu\nu}$. This means that the relation (\ref{ricci}) between the tensors $\R_{\mu\nu}$ and $R_{\mu\nu}$ can be now rewritten as
\begin{equation} \label{eq:conformal_Rmn}
\R_{\mu\nu}=R_{\mu\nu}+\frac{3}{2\phi^2}\partial_\mu \phi \partial_\nu \phi-\frac{1}{\phi}\left(\nabla_\mu
\nabla_\nu \phi+\frac{1}{2}g_{\mu\nu}\nabla_\alpha \nabla^\alpha \phi\right) \ ,
\end{equation}
which can be used in the action (\ref{eq:S_scalar1}) to get rid of the independent connection and to obtain the following
scalar-tensor representation that belongs to the ``Algebraic Family of Scalar-Tensor Theories'' \cite{Koivisto:2009jn}, so that one finally arrives at the following action 
\begin{equation} \label{eq:S_scalar2}
S= \frac{1}{2\kappa^2}\int {\rm d}^4 x \sqrt{-g} \left[ (\OA+\phi)R +\frac{3}{2\phi}\partial_\mu \phi \partial^\mu \phi
-V(\phi)\right] +S_m \ .
\end{equation}
It is interesting to point out that, by the substitution $\phi \rightarrow -(\kappa\phi)^2/6$, the action (\ref{eq:S_scalar2}) reduces to the well-known case of a conformally coupled scalar field with a self-interaction potential. Namely, this redefinition makes the kinetic term in the action (\ref{eq:S_scalar2}) the standard one, and the action itself becomes that of a massive scalar-field conformally coupled to the Einstein gravity.  Of course, it is not the Brans-Dicke gravity where the scalar field is massless but now we have a nonzero $V(\phi)$ as given in (\ref{eq:S_scalar1}). 

In the limit $\OA\rightarrow 0$, the theory (\ref{eq:S_scalar2}) presents the Palatini-$f(\R)$ gravity, and in the limit $\OA\rightarrow \infty$ the metric $f(R)$ gravity \cite{Koivisto:2009jn}. Apart from these singular cases, the more generic theories with a finite $\OA$ thus lie in the ''hybrid'' regime, which from this perspective provides a unique interpolation between the two a priori completely distinct classes of gravity theories. In fact, we have arrived at Brans-Dicke type of theories specified by the non-trivial coupling function
\be
\omega_{BD}=\frac{3\phi}{2\phi-2\Omega_A}\,,
\ee
that generalises the $\omega_{BD}=0$ and $\omega_{BD}=-3/2$ cases, which correspond to the scalar-tensor representations of the metric $f(R)$ and the Palatini-$f(\R)$ gravities \cite{fRgravity6}, respectively. 

Using Eq.~(\ref{eq:conformal_Rmn}) and Eq.~(\ref{eq:var-phi}) in Eq.~(\ref{eq:var-gab}), the metric field equation can be
written as
\begin{eqnarray}
(\OA+\phi) R_{\mu\nu}&=&\kappa^2\left(T_{\mu\nu}-\frac{1}{2}g_{\mu\nu} T\right)+\frac{1}
{2}g_{\mu\nu}\left(V+\nabla_\alpha \nabla^\alpha \phi\right)+\nabla_\mu\nabla_\nu\phi-\frac{3}{2\phi}\partial_\mu \phi
\partial_\nu \phi \ \label{eq:evol-gab} ,
\end{eqnarray}
or equivalently as
\begin{eqnarray}\label{einstein_phi}
(\OA+\phi)G_{\mu\nu}=\kappa^2T_{\mu\nu} + \nabla_\mu\nabla_\nu\phi -
\nabla_\alpha \nabla^\alpha \phi\/g_{\mu\nu}
 -\frac{3}{2\phi}\nabla_\mu\phi \nabla_\nu\phi +
\frac{3}{4\phi}\nabla_\lambda\phi\nabla^\lambda\phi g_{\mu \nu}-
\frac{1}{2}Vg_{\mu\nu},
\end{eqnarray}
from which it is seen that the spacetime curvature is generated by both the matter and the scalar field.
The scalar field equation can be manipulated in two different ways that illustrate further how the hybrid models combine physical features of the $\omega_{BD}=0$ and $\omega_{BD}=-3/2$ scalar-tensor models. 

First, tracing Eq.~(\ref{eq:var-gab}) with $g^{\mu\nu}$,  we find $-\OA R-\phi\R+2V=\kappa^2T$, and using Eq.~(\ref{eq:var-phi}), it takes the following form:
\begin{equation}\label{eq:phi(X)}
2V-\phi V_\phi=\kappa^2T+\OA R \ .
\end{equation}
Similarly as in the Palatini ($\omega_{BD}=-3/2$) case, this equation tells
us that the field $\phi$ can be expressed as an algebraic function
of the scalar $X\equiv \kappa^2T+\OA R$, i.e., $\phi=\phi(X)$. In the
pure Palatini case, however, $\phi$ is just a function of $T$. The
right-hand side of Eq.~(\ref{eq:evol-gab}), therefore, besides
containing new matter terms associated with the trace $T$ and its
derivatives, also contains the curvature $R$ and its derivatives.
Thus, this theory can be seen as a higher-derivative theory in
both  matter and  metric fields. However, such an
interpretation can be avoided if $R$ is replaced in
Eq. (\ref{eq:phi(X)}) with the relation 
\begin{equation}
R=\R+\frac{3}{\phi}\nabla_\mu \nabla^\mu 
\phi-\frac{3}{2\phi^2}\partial_\mu \phi \partial^\mu \phi
\end{equation}
together with $\R=V_\phi$. One then finds that the scalar field is
governed by the second-order evolution equation that becomes, when $\OA=1$,
\begin{equation}\label{eq:evol-phi}
-\nabla_\mu \nabla^\mu \phi+\frac{1}{2\phi}\partial_\mu \phi \partial^\mu
\phi+\frac{\phi[2V-(1+\phi)V_\phi]} {3}=\frac{\phi\kappa^2}{3}T\,,
\end{equation}
which is an effective Klein-Gordon equation.
This last expression shows that, unlike in the Palatini ($\omega_{BD}=-3/2$)
case, the scalar field is dynamical. The theory is therefore not
affected by the microscopic instabilities that arise in Palatini
models with infrared corrections \cite{Olmo:2011uz}.

Finally, we can make a conformal transformation into the Einstein frame of these theories. The conformal rescaling we need to achieve this is given by
\be
\hat{g}_{\mu\nu} \equiv \lp \phi + \OA\rp g_{\mu\nu}\,, 
\ee
and the Einstein frame Lagrangian then becomes
\be \label{einsteinframe}
\hat{\mathcal{L}} = \hat{R} + \frac{3\OA}{2\phi}\frac{\hat{g}^{\al\bt}\phi_{,\al}\phi_{,\bt}}{\lp \phi + \OA\rp^2} -  \frac{V(\phi)}{\lp \phi + \OA\rp^2}\,.
\ee
This can be further put into its canonical form by introducing the rescaled field $\psi$ as
\be
\phi = \OA \tan^2\lp \frac{\psi}{2\sqrt{3}}\rp\,.
\ee
The vacuum theory then becomes a canonical scalar theory with a very specific potential (stemming of course from the original function $f(\R)$) in the Einstein frame.


\subsection{The Cauchy problem}\label{Sec:III}

The dynamical equivalence with scalar-tensor theories shown above is useful to discuss the well--posedness of the Cauchy problem for hybrid $f(X)$-gravity in vacuo and coupled to standard matter sources. For previous studies of the Cauchy problem in different formulations of $f(R)$ theories, see \cite{Cauchy-others1,Cauchy-others1b,Cauchy-others1c,Cauchy-others1d, Cauchy-others2}.

We begin by proving the well--posedness of the Cauchy problem in vacuo, making use of the equivalent formulation (\ref{einstein_phi}) and (\ref{eq:evol-phi}). 
As we shall argue, the same conclusions hold in the presence of standard matter sources satisfying the usual conservation laws $\nabla^\mu\/T_{\mu\nu}=0\/$. 

Borrowing definitions and notations from \cite{yvonne4}, the key point of our discussion is the introduction of suitable generalized harmonic coordinates, defined by the conditions
\begin{equation}\label{2.2.1}
F^\mu_{\phi}:= F^\mu - H^\mu =0 \qquad {\rm with}\qquad F^\mu :=g^{\alpha\beta}\Gamma^\mu_{\alpha\beta}, \quad H^\mu := \frac{1}{(1+\phi)}\nabla^\mu\/\phi\,.
\end{equation}
As we shall see, the gauge (\ref{2.2.1}) allows us to develop a second order analysis very similar to the one used in GR \cite{yvonne4}. 
We notice that the generalized harmonic gauge (\ref{2.2.1}) is a particular case of the one introduced in \cite{Salgado} to prove the well-posedness of the Cauchy problem for a certain class of scalar-tensor theories of gravity.
 
Let us  start with rewriting  Eqs. (\ref{einstein_phi}) in the form \cite{Capozziello:2013gza}
\begin{equation}\label{2.2.2}
R_{\mu\nu} = \frac{1}{(1+\phi)}\left[\Sigma_{\mu\nu} - \frac{1}{2}\Sigma\/g_{\mu\nu}\right]\,,
\end{equation}
where
\begin{equation}\label{2.2.3}
\Sigma_{\mu\nu}:= \nabla_\mu\nabla_\nu\phi - \nabla_\alpha\nabla^\alpha \phi\/g_{\mu\nu} - \frac{3}{2\phi}\nabla_\mu\phi
\nabla_\nu\phi + \frac{3}{4\phi}\nabla_\lambda\phi\nabla^\lambda\phi - \frac{1}{2}Vg_{\mu\nu}  \,,
\end{equation}
plays the role of an effective energy--momentum tensor. We recall that the  Ricci tensor 
can be expressed as \cite{yvonne4}
\begin{equation}\label{2.2.4}
R_{\mu\nu} = R_{\mu\nu}^\phi + \frac{1}{2}\left[ g_{\mu\sigma}\partial_\nu\left( F^\sigma_\phi + H^\sigma \right) + g_{\nu\sigma}\partial_\mu\left( F^\sigma_\phi + H^\sigma \right)\right]\,,
\end{equation}
with
\begin{equation}\label{2.2.5}
R_{\mu\nu}^\phi := - \frac{1}{2}g^{\alpha\beta}\partial^2_{\alpha\beta}\/g_{\mu\nu} + A_{\mu\nu}\/(g,\partial g)\,,
\end{equation}
where only  first order derivatives  appear in the functions $A_{\mu\nu}$. Due to the assumed gauge condition $F^\mu_\phi =0$ and the explicit expression of $H^\mu$, from (\ref{2.2.4}) and (\ref{2.2.5}),  we get the following representation
\begin{equation}\label{2.2.6}
R_{\mu\nu} = - \frac{1}{2}g^{\alpha\beta}\partial^2_{\alpha\beta}\/g_{\mu\nu} + \frac{1}{(1+\phi)}\partial^2_{\mu\nu}\/\phi + B_{\mu\nu}\/(g,\phi,\partial g,\partial\phi)\,,
\end{equation}
where the functions $B_{\mu\nu}$ depend on the metric $g$, the scalar field $\phi$ and their first order derivatives. At the same time, using Eq. (\ref{eq:evol-phi}) to replace all terms depending on the divergence $g^{\alpha\beta}\nabla_\alpha\nabla_\beta\/\phi$, the right hand side of (\ref{2.2.2}) can be expressed as 
\begin{equation}\label{2.2.7}
\frac{1}{(1+\phi)}\left[\Sigma_{\mu\nu} - \frac{1}{2}\Sigma\/g_{\mu\nu}\right] = \frac{1}{(1+\phi)}\partial^2_{\mu\nu}\/\phi + C_{\mu\nu}\/(g,\phi,\partial g,\partial\phi)\,,
\end{equation}
where, again, the functions $C_{\mu\nu}$ depend only on first order derivatives. A direct comparison of Eq. (\ref{2.2.6}) with Eq. (\ref{2.2.7}) shows  that, in the considered gauge, Eq. (\ref{2.2.2}) assumes the form  
\begin{equation}\label{2.2.8}
g^{\alpha\beta}\partial^2_{\alpha\beta}\/g_{\mu\nu} = D_{\mu\nu}\/(g,\phi,\partial g,\partial\phi)\,.
\end{equation}

The conclusion follows that Eq. (\ref{eq:evol-phi}) together with  Eq. (\ref{2.2.8}), form a quasi-diagonal, 
quasi-linear second-order system of partial differential equations, for which well known theorems by Leray \cite{yvonne4,Leray,Wald} hold. Given initial data on a space-like surface, 
the associated Cauchy problem is then well-posed in suitable Sobolev spaces \cite{yvonne4}. 
Of course, the initial data have to satisfy the gauge conditions $F^i_{\phi}=0$ as well as the Hamiltonian and momentum constraints
\begin{equation}\label{2.2.9}
G^{0\mu}=\frac{1}{(1+\phi)}\/\Sigma^{0\mu} \quad \mu=0,\ldots,3  \,,
\end{equation}
on the initial space-like surface. In connection with this, we notice that, from Eq. (\ref{eq:evol-phi}), 
we can derive the expression of the second partial derivative $\partial^2_0\/\phi$ and replace it on the right hand side of (\ref{2.2.9}), and thus obtaining constraints involving no higher than first order partial derivatives with respect to the time variable $x^0\/$. To conclude, we have to prove that the gauge conditions $F^\mu_\phi =0$ are preserved in a neighbourhood of the initial space-like surface. To this end, we first verify that the divergence of the  gravitational field equation (\ref{einstein_phi}) vanishes, namely
\begin{equation}\label{2.2.10}
\nabla^\mu\/\left[(1+\phi)\/G_{\mu\nu} - \Sigma_{\mu\nu}\right] =0 \,.
\end{equation}

Taking into account the identities $\nabla^\mu\/G_{\mu\nu}=0\/$ and $\left(\nabla^\mu\/\phi\right)\/R_{\mu\nu}=\left( \nabla^\mu\nabla_\mu\nabla_\nu - \nabla_\nu\nabla^\mu\nabla_\mu \right)\phi$, automatically satisfied by the Einstein and Ricci tensors, we have
\begin{equation}\label{2.2.11}
\nabla^\mu\/\left[(1+\phi)\/G_{\mu\nu} - \Sigma_{\mu\nu}\right] = -\frac{1}{2}R\nabla_\nu\phi +\nabla^\mu\left( \frac{3}{2\phi}\nabla_\mu\phi\nabla_\nu\phi - \frac{3}{4\phi}\nabla_\lambda\phi\nabla^\lambda\phi\/g_{\mu\nu} + \frac{1}{2}V(\phi)g_{\mu\nu}\right) \,.
\end{equation}
On the other hand, inserting the content of Eq. (\ref{eq:phi(X)}) (in this case, with $T=0$) into the trace of the field equation (\ref{einstein_phi}), we end up with the identity
\be\label{identityR}
R=\frac{dV}{d\phi} + \frac{3}{\phi}\nabla_\lambda\nabla^\lambda\phi - \frac{3}{2\phi^2}\nabla_\lambda\phi\nabla^\lambda\phi \,.
\ee
The identities (\ref{2.2.10}) follow then from a direct comparison of (\ref{2.2.11}) with (\ref{identityR}). 

Now, if $g_{\mu\nu}$ and $\phi$ solve the reduced field Eq. 
(\ref{2.2.8}) and  the scalar field Eq. (\ref{eq:evol-phi}), then we have
\begin{equation}\label{2.2.12}
(1+\phi)\/G^{\mu\nu} - \Sigma^{\mu\nu} = - \frac{(1+\phi)}{2}\/\left( g^{\mu\sigma}\partial_\sigma\/F^\nu_\phi + g^{\nu\sigma}\partial_\sigma\/F^\mu_\phi - g^{\mu\nu}\partial_\sigma\/F^\sigma_\phi \right) \,.
\end{equation}
Identities (\ref{2.2.10}) imply then that the functions $F^\mu_\phi$ satisfy necessarily a linear homogeneous system of wave equations of the form
\begin{equation}\label{2.2.13}
g^{pq}\partial^2_{pq}\/F^i_\varphi + E^{iq}_p\/\partial_q\/F^p_\varphi =0  \,,
\end{equation}
where $E^{iq}_p\/$ are known functions on the space-time. Since the constraints (\ref{2.2.9}) amount to the condition $\partial_0\/F^i_\varphi =0$  on the initial space-like surface, a well known uniqueness theorem for differential systems such as Eq. (\ref{2.2.13}) assures that $F^i_\varphi=0$ in the region where solutions of Eqs. (\ref{eq:evol-phi}) and (\ref{2.2.8}) exist (see also \cite{yvonne4}).

The illustrated analysis also applies in the case of couplings to standard matter sources such as electromagnetic or Yang-Mills fields, (charged) perfect fluid, (charged) dust, Klein-Gordon scalar fields \cite{Cauchy-others1d}, so showing the well-posedness of the Cauchy problem for $f(X)$-gravity in presence of standard matter fields. Indeed, when matter sources are present, Eqs. (\ref{eq:evol-phi}) and (\ref{2.2.8}) have to be coupled with the matter field equations. Applying the same arguments developed for GR \cite{yvonne4,yvonne2,yvonne}, it is easily seen that, in the generalized harmonic gauge (\ref{2.2.1}), the matter field equations together with Eqs. (\ref{eq:evol-phi}) and (\ref{2.2.8}) form a Leray hyperbolic and a causal differential system  admitting a well-posed Cauchy problem \cite{Leray}. In addition to the well-known results by Bruhat, the crucial  point is again that the field equations of matter field imply the standard conservation laws $\nabla^\mu\/T_{\mu\nu}=0$ \cite{Koivisto:2005yk}. 
In summary, the hybrid metric-Palatini gravity satisfies the  well-formulation and well-posedness of Cauchy problem for standard forms of matter and then, in this sense, it is a viable theory.

\subsection{More general hybrid metric-Palatini theories}
\label{tts}

The ``hybrid'' theory space is a priori large. In addition to the metric and its Levi-Civita connection, one also has an additional independent connection as a building block to construct curvature invariants from. Thus one can consider various new terms such as
\begin{align}
\hat{R}^{\mu\nu}\hat{R}_{\mu\nu}\,, \quad R^{\mu\nu}\hat{R}_{\mu\nu}\,, \quad \hat{R}^{\mu\nu\al\bt}\hat{R}_{\mu\nu\al\bt}\,, \quad \quad R^{\mu\nu\al\bt}\hat{R}_{\mu\nu\al\bt}\,, \quad \R R\,, \quad \text{etc}\,.
\end{align}
Though an exhaustive analysis of such hybrid theories has not been performed, there is some evidence that the so called $f(X)$ class of theories we are focusing our attention upon here is a unique class of viable higher order hybrid gravity theories. In the more restricted framework of purely metric theories, it is well known that the $f(R)$ class of theories is exceptional by avoiding the otherwise generic Ostrogradski instabilities by allowing a separation of the additional degrees of freedom into a harmless scalar degree of freedom \cite{Woodard:2006nt}: as we have already seen, such a separation is possible also for our hybrid-$f(X)$ theories. Furthermore, it turns out that this feature is a similar exception in the larger space of metric-affine theories, since a generic theory there is inhabited by ghosts, superluminalities or other unphysical degrees of freedom. 

As a representative class of more general theories, actions of the form  
\begin{align}
S = \frac{1}{2\kappa^2} \int {\rm d}^4x \sqrt{-g}\, f( R,\R,\hat Q_H )\,, \qquad \hat Q_H = R^{\mu\nu}\hat{R}_{\mu\nu} 
\label{001}
\end{align}
were studied by Tamanini \cite{TT}, who determined the precise field content of this action in the weak-field limit.
Variation of (\ref{001}) with respect to the metric produces the following field equations:
\bea
f_{,R} R_{\mu\nu} -\frac{1}{2}g_{\mu\nu}f +g_{\mu\nu}\Box f_{,R} -\nabla_\mu\nabla_\nu f_{,R} +f_{,\R} \hat{R}_{\mu\nu} 
 +  2f_{,\hat{Q}}R_\mu^\lambda \hat{R}_{\nu\lambda} && \nonumber \\ +\frac{1}{2}\Box \left(f_{,\hat{Q}}\hat{R}_{\mu\nu}\right) +\frac{1}{2}g_{\mu\nu}\nabla_\alpha\nabla_\beta\left(f_{,\hat{Q}} \hat{R}^{\alpha\beta}\right) -\nabla_\lambda\nabla_{(\nu}\left(f_{,\hat{Q}} \hat{R}_{\mu)}^{\lambda}\right) & = & \kappa^2 T_{\mu\nu} \,,
\label{003}
\eea
where $f_{,R}$, $f_{,\R}$ and $f_{,\hat{Q}}$ are the derivatives of $f$ with respect to $R$, $\R$ and $\hat Q_H$ respectively. The solution to the equation of motion for the
connection on the other hand dictates that it is the Levi-Civita connection of the metric $\hat{g}_{\mu\nu}$ given by
\be
\hat g^{\mu\nu} = \frac{\sqrt{-r}}{\sqrt{-g}}\,r^{\mu\nu} \,, \quad \text{where} \quad r^{\mu\nu} = f_{,\R}g^{\mu\nu} +f_{,\hat{Q}}R^{\mu\nu} \,.
\ee
Using this one can eliminate the auxiliary metric $\hat{g}_{\mu\nu}$ in terms of the physical metric $g_{\mu\nu}$. 

Considering perturbations $h_{\mu\nu}=g_{\mu\nu}-\eta_{\mu\nu}$ around Minkowski space $g_{\mu\nu}=\eta_{\mu\nu}$, and
inverting the linearised field equations for the physical metric then gives us the propagators for the graviton and the additional degrees of freedom that may be present in $h_{\mu\nu}$. 
The propagator $\Pi^{\al\bt\gamma\delta}$ is defined by
\be
\Pi^{-1\gamma\delta}_{\al\bt}h_{\gamma\delta}=\kappa^2 \tau_{\al\bt}\,,
\ee
where $\tau_{\al\bt}$ represents the linearised stress energy source. In the formalism of the spin-projector operators employed in Ref. \cite{spin1} and more pedagogically reviewed in Ref. \cite{spin2}, the result can be given in Fourier space (where basically $\Box \rightarrow -k^2$) in terms of two functions $a$ and $c$ as
\be
k^2\Pi_{\al\bt\gamma\delta} =  \frac{\mathcal{P}^2_{\al\bt\gamma\delta}}{a(-k^2)} - \frac{\mathcal{P}^0_{\al\bt\gamma\delta}}{ a(-k^2)-3c(-k^2)}\,,
\ee
where $\mathcal{P}^2_{\al\bt\gamma\delta}$ picks up the spin-2 and $\mathcal{P}^0_{\al\bt\gamma\delta}$ the scalar modes of the fluctuations, see Refs. \cite{TT,spin1,spin2} for details. The functions $a$ and $c$ can be determined immediately given a theory of the form (\ref{001}). They depend upon the combinations
\begin{align}
A = \frac{6f^{(0)}_{\R \R} +f^{(0)}_{,\hat Q}}{2f^{(0)}_{,\R}} \,, \quad\mbox{and}\quad B=\frac{f^{(0)}_{,\hat Q}}{f^{(0)}_{\R}} \,,
\end{align}
in the following way:
\bea
a(\Box) & = & f^{(0)}_{,R}+f^{(0)}_{,\R} -f^{(0)}_{,\hat Q}\frac{B}{4}\Box^2 \,, \label{abox} \\
c(\Box) &= & f^{(0)}_{,R}+f^{(0)}_{,\R} -2\left(f^{(0)}_{,RR}+4f^{(0)}_{,R\R}+f^{(0)}_{,\hat Q}\right)\Box +\left[f^{(0)}_{,R\R}\left(6A+B\right) +f^{(0)}_{,\hat Q}\left(2A+\frac{B}{4}\right)\right]\Box^2 \,. \\
\eea
Let us then enumerate some special cases. To simplify things we assume $f^{(0)}_{\R\R}=0$.

\subsubsection{Metric $f(R)$ models}

In the pure metric $f(R)$ case, $\f=A=0$ and we have
\be \label{scalaron}
\Pi_{f(R)}^{\al\bt\gamma\delta} = \Pi_{GR}^{\al\bt\gamma\delta} + \frac{1}{2\lp k^2+ (3\Frr)^{-1}\rp}\mathcal{P}^{0\al\bt\gamma\delta} \,.
\ee
Thus we have an extra scalar degree of freedom, as we expect since the $f(R)$ models are known to be equivalent to Brans-Dicke theories with a vanishing parameter $\omega_{BD}=0$. 
The mass of the ``scalaron'' is $m^2= (3\Frr)^{-1}$, and as long as $f''(R)>0$ the theory is stable, otherwise a tachyonic mass spoils the stability around Minkowski space. 

\subsubsection{Palatini $f(\R)$ models}

As already discussed, the Palatini-type $f(\R)$ models are equivalent to Brans-Dicke theories with the parameter $\omega_{BD}=-3/2$. This particular value corresponds to vanishing kinetic term of the field, which thus is nondynamical.  Therefore we expect that no additional scalar degree of freedom should appear. For a proper normalisation we may assume that $f^{(0)}_{,\R}=1$, and we have now of course that $\Frr=\f=f^{(0)}_{,\hat Q}=0$. Hence,
\be
\Pi_{f(\R)}^{\al\bt\gamma\delta} =\Pi_{GR}^{\al\bt\gamma\delta} \,,
\label{013}
\ee
confirming our expectation.

\subsubsection{Hybrid $f(X)$ models}

It was already remarked in \cite{Koivisto:2009jn}, in Ricci-flat spacetimes the $f(X)$ theories share the properties of Palatini-$f(\R)$ theories, which in vacuum reduce to GR with a possible cosmological constant. 
Therefore it is not a surprise that we find no new propagating degrees of freedom in Minkowski vacuum,
\be
\Pi_{f(X)}^{\al\bt\gamma\delta}=\Pi_{GR}^{\al\bt\gamma\delta}\,.
\label{016}
\ee
 Interestingly though, this class of theories is not equivalent to either of the previous two cases, since when one considers curved spacetimes, a new scalar degree of freedom appears. In this sense, the $f(X)$ gravity is a quite minimalistic scalar-tensor extension of GR, as the scalar propagates only in the presence of background curvature.

\subsubsection{The hybrid $f(R,\R)$ models}

The generalized hybrid Ricci scalar theories were introduced in \cite{Flanagan:2003iw,Bo} and found to have qualitatively different properties compared to the more restricted class of $f(X)$ models described above. In particular, the $f(R,\R)$ were shown to be equivalent to a class of biscalar-tensor theories. These theories have an extra $\mathcal{P}^{0\al\bt\gamma\delta}$ spin-0 propagator with a double pole, corresponding to two propagating scalar degrees of freedom. We can easily deduce the masses of these scalar fields. We get
\be
m^2_\pm = \frac{f^{(0)}_{,\R}}{18\lp \f\rp^2}\lp \Frr+4\f \pm S\rp\,,
\ee
where we have defined for convenience
\be
S  \quad \equiv \sqrt{\lp\Frr+4\f\rp^2- 12\frac{\lp\f\rp^2}{f^{(0)}_{,\R}} }\,.
\ee
We note that the scalar particle with mass squared $m_-^2$ corresponds to the scalaron appearing in (\ref{scalaron}) in the limit of pure $f(R)$ gravity, but in general now has a shifted mass. The other scalar is a new particle that occurs due to nontrivial dependence upon $\R$, and unlike in the case of $f(X)$ gravity, it propagates also in Ricci-flat spaces. The condition that neither of the scalars has a tachyonic instability, is given by
\be \label{frrc}
f^{(0)}_{,\R}  >  0\,, \quad \mbox{and} \quad
\Frr+4\f-S> 0\,.
\ee
The residues at the two poles corresponding to these masses are
\ba
r_\pm = \frac{S \pm \lp\Frr+ 4\f \rp}{4S}\,.
\ea
In order for neither of these scalars to be a ghost, we should have both $r_+>0$ and $r_->0$. The second condition would require that
\be
\Frr+4\f-S < 0\,,
\ee
in contradiction with (\ref{frrc}). It seems then that we cannot avoid both tachyons and ghosts in this theory.

\subsubsection{The hybrid Ricci-squared $f(\R,\hat{Q})$ theories}

Let us finally consider the $\hat{Q}_H$-invariant. For simplicity, we restrict to models here without nonlinear dependence on the metric Ricci scalar; it is easy to see that this does not affect our conclusions essentially. 
Basically the graviton propagator acquires its structure from the function $a(\Box)$ in (\ref{abox}), and now only the higher-derivative term $\hat{Q}_H$ modifies it. We can arrange the result for the propagator in the form 
\begin{align}
\Pi_{f(\hat{R},\hat{Q})}^{\al\bt\gamma\delta} = \frac{\Pi_{GR}^{\al\bt\gamma\delta}}{\lp 1-\frac{1}{4}\lp\Q\rp^2 k^4\rp} +\frac{3\Q\lp 1+\frac{3}{4}\Q k^2\rp}{2\lp 1-\frac{1}{4}\lp\Q\rp^2k^4\rp\lp 1+3\Q k^2+2\lp\Q\rp^2 k^4\rp}\mathcal{P}^{0\al\bt\gamma\delta} \,.
\end{align}
The sixth order theory we have at hand has a modulated graviton propagator which adds two extra poles. In addition, there appears a scalar propagator that has five poles. This is in a quite drastic contrast with respect to the metric $Q$-theory which contains only one additional spin-2 particle and features fourth order field equations. We need not analyze in detail the properties of the new degrees of freedom here, since it is obvious the theory as such is seriously haunted by ghosts and thus not physical. It is easy to convince oneself that this occurs very generically once one builds the action from any hybrid curvature invariant -- with the exception of $\R$ in the specific case of separable functional dependence $R + f(\R)$.

These considerations corroborate our claim that the $f(X)$ theories are of special theoretical interest. In the rest of the review, we shall discuss their phenomenology.

\section{Hybrid-gravity cosmology}
 
In order to study the cosmology of the metric-Palatini theories, we choose in this section to employ the scalar-tensor formulation derived above (\ref{eq:S_scalar2}):
\be \label{st2}
S=\frac{1}{2\kappa^2}\int {\rm d}^4 x \sqrt{-g}\lb (\OA+\phi) R + \frac{3}{2\phi}\lp \partial\phi\rp^2 - 2\kappa^2V(\phi)\rb + S_m\,,
\ee
where
\be
\kappa^2V(\phi)=\frac{1}{2}\lb r(\phi)\phi-f(r(\phi))\rb, \quad r(\phi) \equiv {f'}^{-1}(\phi) \,.
\ee
Cosmology of the equivalent theories has been investigated also using the $f(X)$ formulation in terms of the purely metric quantities, and we refer the reader to the recent phase space analysis for the most complete global analysis of the cosmological dynamics of these theories \cite{Carloni:2015bua}. Here we will first write down the cosmological equations in the formulation (\ref{st2}) and then have a brief look at the phase space of exact solutions for these equations. Then we will analyse the formation of cosmological large-scale structure in these models.

\subsection{Background expansion}

The flat Friedmann-Robertson-Walker metric is defined as
\be
{\rm d}s^2 = -{\rm d}t^2 + a^2(t)\lp {\rm d} x^2 +  {\rm d} y^2 +  {\rm d} z^2\rp\,,  
\ee
where the rate of time-evolution of the scale factor $a(t)$ is conveniently parameterised by the Hubble rate $H(t)=({\rm d}a(t)/{\rm d}t)/a(t)$. In the following we will not write explicitly the time dependence of the cosmological background quantities, and denote time derivative by an overdot. This eases the notation and we can write for example $H=\dot{a}/a$. In the following we will mainly be interested in accelerating dark energy -like dynamics; for a study of Einstein static spaces, see Ref. \cite{Boehmer:2013oxa}.

\subsubsection{The Friedmann equations}

The Friedmann equations that govern the evolution of $H$ can always be written in terms of the effective energy density and pressure, respectively defined as
\ba
3H^2 & = & \kappa^2\rho_{\rm eff}\,, \label{fr1} \\
\dot{H} & = & -\frac{\kappa^2}{2}\lp\rho_{\rm eff}+p_{\rm eff}\rp\label{fr2}\,,
\ea
where for the theory defined by the action (\ref{st2}) we obtain the following effective source terms
\ba
\lp \OA+\phi \rp \kappa^2\rho_{\rm eff} & = & -\frac{3}{4\phi}\dot{\phi}^2 + \kappa^2V(\phi) - 3H\dot{\phi} + \kappa^2\rho_m\,, \\
\lp \OA+\phi \rp \kappa^2 p_{\rm eff} & = & -\frac{3}{4\phi}\dot{\phi}^2 - \kappa^2V(\phi) + \ddot{\phi} + 2H\dot{\phi} + \kappa^2 p_m\,,
\ea
respectively.
The conservation equations for the matter component and the scalar field are
\ba
\dot{\rho}_m + 3H(\rho_m + p_m) & = & 0\,, \\
\ddot{\phi} + 3H\dot{\phi} - \frac{\dot{\phi}^2}{2\phi} + \frac{1}{3}\phi R - \frac{2}{3}\kappa^2 \phi V'(\phi) & = & 0 \label{kg}\,.
\ea
Recalling that $R=6(2H^2+\dot{H})$ and using Eqs. (\ref{fr1}) and (\ref{fr2}), we can rewrite the Klein-Gordon equation as
\be
\ddot{\phi}+3H\dot{\phi} - \frac{\dot{\phi}^2}{2\phi} + U'(\phi) + \frac{\kappa^2 \phi}{3\OA}\lp \rho_m-3p_m\rp =0\,, \label{kg2}
\ee
where for notational simplicity, $U'(\phi)$ is defined by
\be
U'(\phi) \equiv \frac{2\kappa^2\phi}{3\OA}\lb 2V(\phi) - \lp \OA + \phi \rp V'(\phi)\rb\,. \label{veff}
\ee
As a consistency check one can verify that the Klein-Gordon equation together with the matter conservation allows to derive (\ref{fr2}) from (\ref{fr1}).
By combining Eqs. (\ref{kg}) and (\ref{kg2}), we find that
\be
2V(\phi)-V'(\phi)\phi =\frac{1}{2}\lp \OA R + \kappa^2 T_m \rp \equiv \frac{1}{2}X\,. \label{alg}
\ee
The solution for $\phi=\phi(X=0)$ gives us the natural initial condition for the field in the early universe. The asymptotic value of the field in the far future may then be deduced by studying the minima of the function $U(\phi)$ defined by Eq. (\ref{veff}).

\subsubsection{Dynamical system analysis}

Cosmological dynamics can be addressed by taking into account a suitable  dynamical system. Let us  introduce the dimensionless variables
\be
\Omega_m  \equiv  \frac{\kappa^2\rho_m}{3H^2}\,, \quad
x  \equiv  \phi\,, \quad y = x_{,N}\,, \quad z = \frac{\kappa^2V}{3H^2}\,,
\ee
where $N=\log{a}$ is the e-folding time. The Friedmann equation (\ref{fr1}) can then be rewritten as
\be
\OA + x + y -z +\frac{y^2}{4x} = \Omega_m\,.
\ee
Due to this constraint, the number of independent degrees of freedom is three instead of four. We choose to span our phase space by the triplet $\{x,y,z\}$. The autonomous system of equations for them reads as
\ba
x_{,N} & = & y\,, \\
y_{,N} & = & \frac{2x+y}{8\OA x}\Big\{ \lp 3w_m-1 \rp y^2 + 4x\lb \lp 3w_m-1 \rp y-3\lp 1+w_m\rp z\rb \nonumber \\
&& - 4x^2\lp 1-3w_m-2u(x)z\rp +4 \OA\lb 3x\lp w_m-1\rp y+y^2-x^2\lp 2-6w_m-4u(x)z\rp\rb\Big\}\,, \\
z_{,N} & = & \frac{z}{4\OA x}\Big\{\lp 3w_m-1\rp y^2  +  4x\lb\lp 3w_m-1\rp y - 3\lp 1+w_m\rp z\rb
\nonumber \\ 
&& + 4\OA x \lp 3+3w_m+u(x)y\rp + 4x^2\lp 3w_m-1+2u(x)z\rp\Big\}\,.
\ea
We have defined $u(x) \equiv V'(\phi)/V(\phi)$, that becomes a constant iff the potential is exponential. The relevant fixed points appear in this system. In particular, we have the matter dominated fixed point where $x=y=z=0$ and $w_{\rm eff}=w_m$, and the de Sitter fixed point\footnote{In addition, there exists the fixed point $x=-\OA$ corresponding to some kind of singular evolution.} that is described by $w_{\rm eff}=-1$ and
\be
x_*=(2-\OA u_*)/u*\,, \quad y_*=0\,, \quad z_* = 2/u_*\,. \label{dS}
\ee
We denote the asymptotic values corresponding to this fixed point by a subscript star. In particular, the asymptotic value of the field $x_*$ is solved from the first equation in (\ref{dS}) once the form of the potential is given. As expected, this value corresponds to minimum of the effective potential (\ref{veff}), $U'(x_*)=0$. To construct a viable model, the potential should be such that we meet the two requirements:
\begin{itemize}
\item The matter dominated fixed point should be a saddle point, the de Sitter fixed point an attractor. Then we naturally obtain a transition to acceleration following standard cosmological evolution.
\item At the present epoch the field value should be sufficiently close to zero. Then we avoid conflict with the Solar System tests of gravity (this will be clarified in Section \ref{weakfield}).
\end{itemize}
Note that the simplest metric $f(R)$ theories that provide acceleration fail in both predicting a viable structure formation era and the Solar system as we observe it. The Palatini-$f(\R)$ models on the other hand can be ruled out as dark energy alternative by considering their structure formation or implications to microphysics, if such a theory is regarded consistent in the first place. As shown here and explored further below, $f(X)$ gravity models exist that are free of these problems.
\newline
\newline
To summarize: the field goes from $\phi_i$ to $\phi_*$, where the former is given by $2V(\phi_i)=V'(\phi_i)\phi_i$ and the latter by $2V(\phi_*)=(\OA+\phi_*)V'(\phi_*)$. We just need a suitable function $V(\phi)$, i.e. $f(\R)$ in such a way that the slope will be downwards and $\phi_*$ near the origin. 

We refer the reader to \cite{Carloni:2015bua} for a more complete and detailed phase space analysis of the cosmological background dynamics.

\subsubsection{On cosmological solutions}

As a specific simple example, let us consider in more detail the specific case of de Sitter solution in vacuum when $\OA=1$. Then the modified
Friedmann equations take the form 
\begin{eqnarray}
3H^2&=& \frac{1}{1+\phi }\left[\kappa^2\rho
+\frac{V}{2}-3\dot{\phi}\left(H+\frac{\dot{\phi}}
{4\phi}\right)\right] \,,\label{field1d} \\
2\dot{H}&=&\frac{1}{1+\phi }\left[
-\kappa^2(\rho_m+p_m)+H\dot{\phi}+\frac{3}
{2}\frac{\dot{\phi}^2}{\phi}-\ddot{\phi}\right] \,,\label{field2d}
\end{eqnarray}
and the scalar field equation (\ref{eq:evol-phi}) becomes
\begin{equation}
\ddot{\phi}+3H\dot{\phi}-\frac{\dot{\phi}^2}{2\phi}+\frac{\phi}{3}
[2V-(1+\phi)V_\phi]=-\frac{\phi\kappa^2}{3}(\rho_m-3p_m) .  \label{3d}
\end{equation}
To further specify the set-up, consider a model that arises by demanding
that matter and curvature satisfy the same relation as in GR.
Taking
\begin{equation} \label{pot1d}
V(\phi)=V_0+V_1\phi^2\,,
\end{equation}
the trace equation automatically implies $R=-\kappa^2T+2V_0$
\cite{Harko:2011nh,Capozziello:2012ny}. As $T\to 0$ with the
cosmic expansion, this model naturally evolves into a de Sitter
phase,  which requires $V_0\sim \Lambda$ for consistency with
observations. If $V_1$ is positive, the de Sitter regime
represents the minimum of the potential.  The effective mass for
local experiments, $m_\varphi^2=2(V_0-2 V_1 \phi)/3$, is then
positive and small as long as $\phi<V_0/V_1$. For sufficiently
large $V_1$ one can make the field amplitude small enough to be in
agreement with Solar System tests. It is interesting that the
exact de Sitter solution is compatible with dynamics of the scalar
field in this model.

The accelerating dynamics that drive the $f(X)$ theory towards its general relativistic limits today have indeed been realised in several specific models \cite{Harko:2011nh,Capozziello:2012ny,Lima:2014aza,Lima:2015nma}. Our preliminary phase space analysis confirmed the existence of de Sitter attractor solutions, and the recent study of cosmology in terms of dynamical system analysis extends this result to more general models \cite{Carloni:2015bua}. Analytic solutions were presented also in Ref. \cite{Capozziello:2012ny} as well as in Ref. \cite{Borowiec:2014wva}, there using a N\"other symmetry technique. A designer approach was deviced by Lima \cite{Lima:2014aza} to reconstruct precisely the standard $\Lambda$CDM expansion history by a nontrivial $f(X)$ model, and finally, two families of models were constrained by confronting their predictions with a combination of cosmic microwave background, supernovae Ia and baryonic acoustic oscillations background data \cite{Lima:2015nma}.


\subsection{Cosmological perturbations}\label{sec:pert}

To understand the implications of these models to the cosmological structure formation, we will derive the perturbation equations and analyse them in some
specific cases of interest. This paves the way for a detailed comparison of the predictions with the cosmological data on large scale structure and the cosmic microwave background. For generality, we will keep the parameter $\Omega_A$ in the formulas in this section.



\subsubsection{Field equations and conservation laws}

We work in the Newtonian gauge \cite{Ma:1995ey}, which can be parameterized by the two gravitational potentials $\Phi$ and $\Psi$,
\be \label{newtongauge}
{\rm d}s^2= -\lp 1+2\Psi \rp {\rm d}t^2+a^2(t)\lp 1+2\Phi\rp {\rm d}\vec{x}^2\,. 
\ee
As matter source we consider a perfect fluid, with the background equation of state $w$ and with density perturbation $\delta = \delta\rho_m/\rho_m$, pressure perturbation $\delta p_m=c_s^2 \delta\rho_m$ and velocity perturbation $v$.

The $0$-$0$ part of the field equations is
\ba
\frac{k^2}{a^2}\Phi 
+ 3\lp H-\frac{\dot{\phi}}{2\Fp}\rp \dot{\Phi} 
- 3\lp H^2 +\frac{H\dot{\phi}}{\F}-\frac{\dot{\phi}^2}{4\phi \Fp}\rp\Psi   
   \nonumber \\ 
=\frac{1}{2\F}\lb \kappa^2\delta\rho_m + \lp \frac{3}{4\phi^2}\dot{\phi}^2+V'(\phi)-3H^2-\frac{k^2}{a^2}\rp\varphi - 3\lp H + \frac{\dot{\phi}}{2\phi}\rp 
\dot{\varphi}\rb\,,
\ea
where we have denoted $\varphi=\delta\phi$.
The Raychaudhuri equation for the perturbations reads
\ba
\lb 6\lp H^2+2\dot{H}\rp - 2\frac{k^2}{a^2} + \frac{6}{\F}\lp 
\ddot{\phi}-\frac{\dot{\phi}^2}{\phi^2}+H\dot{\phi}\rp \rb \Psi 
+ 3\lp 2H - \frac{\dot{\phi}}{\F}\rp \lp \dot{\Phi} - \dot{\Psi}\rp 
-6\ddot{\Phi} =
 \nonumber \\
 \frac{1}{\F}\lb \kappa^2\lp \delta\rho_m+3\delta p_m\rp 
+\lp 6H^2+6\dot{H} + 3\frac{\ddot{\phi}}{\phi^2}-2V'(\phi)+\frac{k^2}{a^2}\rp\varphi
+3\lp H-\frac{2\dot{\phi}}{\phi}\rp \dot{\varphi}
+ 3\ddot{\varphi}
\rb .
\ea
The $0$-$i$ equation is
\be
- \lp H+\frac{\dot{\phi}}{2\Fp}\rp\Phi + \dot{\Phi}
= \frac{1}{2\Fp}\lb \kappa^2\lp\rho_m+p_m\rp a v_m + \lp H+\frac{3\dot{\phi}}{2\phi}\rp\varphi+\dot{\varphi}\rb\,.
\ee
Note that the set of perturbed field equations is completed by the off-diagonal spatial piece:
\be
\Psi+\Phi=-\frac{\varphi}{\F}\,.
\ee
Assuming a perfect fluid, the continuity and Euler equations for the matter component are 
\ba
\dot{\delta} + 3H\lp c_s^2-w\rp\delta & = & -\lp 1+w \rp \lp 3\dot{\Phi}-\frac{k^2}{a}v\rp\,, \\
\ddot{v}+\lp 1-3c_a^2\rp Hv & = &  \frac{1}{a}\lp \Psi + \frac{c_s^2}{1+w}\delta\rp\,, 
\ea
respectively. The linear part of the Klein-Gordon equation is then compatible with the above system. For completeness, it is 
\ba
\ddot{\varphi}+\lp 3H+\frac{1}{\phi}\rp \dot{\varphi} + \lp \frac{k^2}{a^2}+\frac{\dot{\phi}^2}{2\phi^2}- \frac{2}{3}V''(\phi)\rp\varphi &= & \nonumber
\lp 2\ddot{\phi}+6H\dot{\phi}-\frac{3}{2\phi}\dot{\phi}^2\rp\Psi + \dot{\phi}\lp \dot{\Psi}-3\dot{\Phi}\rp - \frac{\phi}{3}\delta R\,.
\ea
This completes the presentation of the field equations and the conservation laws. For the equations in the synchronous gauge, see \cite{Lima:2014aza}.

\subsubsection{Matter dominated cosmology}

Let us consider the formation of structure in the matter-dominated universe, where $w=c_s^2=0$. 
In this subsection, we shall consider scales deep inside the Hubble radius. This so called quasi-static approximation is well known in the literature and indeed we will arrive at a similar result as have been known to apply for scalar-tensor theories since early studies \cite{Boisseau:2000pr} that have been more recently generalised to wide a variety of coupled dark sector models \cite{coupled}. The approximation neglects a fluctuating degree of freedom that is expected to be insignificant at small enough scales. For analyses of the applicability of the approximation with different assumptions on the cosmological models, see \cite{coupled, delaCruzDombriz:2008cp,Abebe:2011ry,Llinares:2013qbh,Sawicki:2015zya}. 
  
In the quasi-static subhorison limit the spatial gradients are more important than the time derivatives and, consequently, the 
matter density perturbations are much stronger than the gravitational potentials. Combining the continuity and the Euler equation in this approximation, one obtains
\be
\ddot{\delta}=-2H\dot{\delta}-\frac{k^2}{a^2}\Psi\,.
\ee  
We need then to solve the gravitational potential. Let us define $\Pi = a^2\rho_m\delta/k^2$ and write the field equations and the Klein-Gordon equation at this limit in a very simple way as
\ba
\Fp\Phi &=& \Pi-\varphi\,, \\
\Fp\lp\Psi+\Phi\rp & = & -\varphi\,, \\
-2\Fp\Psi & = & \Pi + \varphi\,, \\
3\varphi & = & -2\phi\lp \Psi + 2\Phi\rp\,.
\ea
We immediately see that one of the equations is (as expected) redundant, and that the $\Psi$ is (as usual) proportional to $\Pi$, where now the proportionality is given as a function 
of the field $\phi$. Our result is
\be \label{delta_evol}
\ddot{\delta} + 2H\dot{\delta} = 4\pi G_{\rm eff}\rho_m\delta\,,
\ee
with
\be \label{delta_evol2}
 G_{\rm eff} \equiv \frac{\OA-\frac{1}{3}\phi}{\OA\Fp}G\,. 
\ee
This shows that instabilities can be avoided in the evolution of the matter inhomogeneities, in contrast to the Palatini-$f(\R)$ models and some matter-coupled scalar field models (recall our theory can be mapped into such in the Einstein frame). Equation (\ref{delta_evol}) provides a very simple approximation to track the growth of structure accurately within the linear regime during matter dominated cosmology. Confrontations of specific model predictions with the present large scale structure data and forecasts for the constraints from future experiments, in particular the Euclid mission \cite{euclid}, is interesting work to be done.

\subsubsection{Vacuum fluctuations}

The propagation of our scalar degree of freedom in vacuum is also a crucial consistency check on the theory. Let us set $\rho_m=0$. Let us consider the curvature perturbation in the uniform-field 
gauge $\zeta$. In terms of the Newtonian gauge perturbations this is 
\be
\zeta= \Phi-\frac{H}{\dot{\phi}}\varphi\,.
\ee
After somewhat more tedious algebra than in the previous case, we obtain the exact (linear) evolution equation
\be \label{eta_evol}
\ddot{\zeta}+\lb 3H
-2\frac{\ddot{\phi}+2\dot{H}\Fp -\frac{\dot{\phi}^2}{\F}}{\dot{\phi}+2H\Fp} 
+ \frac{\phi\Fp}{\dot{\phi}^2}\lp\frac{2\ddot{\phi}\dot{\phi}}{\phi\Fp}
 + \frac{\dot{\phi}^3\Fp^2\phi}{1-\phi^3\Fp^3}\rp\rb\dot{\zeta} = - \frac{k^2}{a^2}\zeta\,.
\ee
The friction term depends on the perturbation variable we consider, but the perturbations at small scales still propagate with the speed of light, as in canonical scalar field 
theory. This excludes also gradient and tachyon instabilities in the graviscalar sector. Now equation (\ref{eta_evol}) can be used to study 
generation of fluctuations in $f(X)$-inflation. Construction of specific models and their observational tests are left for forthcoming studies; the Einstein-frame formulation (\ref{einsteinframe}) might present a convenient starting point for that as it, given the function $f(\R)$, presents directly the relevant inflationary potential in terms of the canonic field.

\section{Astrophysical applications} \label{astro}

In this Section we develop methods to study dark matter phenomenology in the hybrid models; for studies of dark matter generated by metric $f(R)$ modification, 
see e.g. \cite{Cembranos:2008gj,Arbuzova:2011fu}.

Hybrid gravity allows to address several issues related to dark matter dynamics ranging from galaxies to galaxy clusters. 
These self-gravitating structures can be probed by studying the motion of test particles (stars for galaxies and galaxies themselves for galaxy clusters) moving into a gravitational potential. The behaviour of  rotational and dispersion velocities of such test particles  can be explained within the framework of the gravitational potential derived from the theory.   For example, the tangential velocity can be explicitly obtained as a function of the scalar field of the equivalent scalar-tensor description. The model predictions can be  compared  with samples of rotation curves of spiral galaxies and galaxy clusters, respectively.  
The possibility of constraining the form of the scalar field and the parameters of the model by using the stellar velocity dispersions is also analysed. 
Furthermore, the Doppler velocity shifts are also obtained in terms of the scalar field. Finally suitable generalizations of the virial theorem and the relativistic Boltzmann equation allow to construct a self-consistent theory for galaxy clusters.
In conclusion, all the physical and geometrical quantities and the numerical parameters in the hybrid metric-Palatini model can be expressed in terms of observable/measurable quantities, such as the tangential velocity, the baryonic mass of the galaxy, the Doppler frequency shifts,  the  dispersion velocity, the geometrical quantities characterizing the clusters of galaxies respectively. These results open the possibility of testing the hybrid metric-Palatini gravitational models  at the galactic or extra-galactic scale by using direct astronomical and astrophysical observations. 

Let us start our considerations by dealing with the weak field limit of the theory.

\subsection{The weak field limit}
\label{weakfield}

It is of paramount importance to determine the post-Newtonian parameters of the theory as they determine the compatibility of the theory with
the local precision gravity tests. For post-Newtonian analysis of the metric and Palatini $f(R)$ theories, see e.g. Refs. \cite{Olmo:2005zr,Olmo:2005hc}, and for a unified analysis Ref. \cite{Koivisto:2011tp}.
Here, in particular, we are interested in the parameter $\gamma$ that is basically the fractional difference of the Newtonian potentials in (\ref{newtongauge}) in the limit where the cosmological expansion can be neglected, $a=1$.

To this end, 
we need to consider the perturbations of Eqs. (\ref{eq:evol-gab}) and (\ref{eq:evol-phi}) in a Minkowskian background. The usual
procedure is to assume $\phi=\phi_0+\varphi(x)$, where $\phi_0$ is
the asymptotic value of the field far away from the local system
(and should be given by the cosmological background solution), and
to take a quasi-Minkowskian coordinate system in which
$g_{\mu\nu}\approx \eta_{\mu\nu}+h_{\mu\nu}$, with
$|h_{\mu\nu}|\ll 1$. Then it is easy to see that the quadratic
terms $\partial_\mu \phi \partial_\nu \phi$ and $\partial_\mu \phi
\partial^\mu \phi$ in Eqs. (\ref{eq:evol-gab}) and (\ref{eq:evol-phi}) do not contribute to the linear order. The
potential terms in Eq.~(\ref{eq:evol-phi}) can be linearized as
follows (in this subsection we denote $V_\phi=V'(\phi)$):
\begin{eqnarray}
\frac{\phi}{3}[2V-(1+\phi)V_\phi]\approx
\frac{\phi_0}{3}[2V_0-(1+\phi_0)V_{\phi_0}]
-\frac{\varphi}{3}\left[\phi_0(1+\phi_0)V_{\phi\phi}|_0+V_{\phi_0}-2V_0\right]\,.
\end{eqnarray}
The zeroth-order term in this equation is due to the background,
and can be absorbed into a coordinate redefinition. The
coefficient of the first-order term can be interpreted as a mass
squared. The linearized scalar field equation is thus given by
\begin{equation}\label{eq:linear-phi}
(\vec{\nabla}^2-m_\varphi^2)\varphi=-\frac{\phi_0\kappa^2}{3}\rho\,,
\end{equation}
where, as usual, in this order of approximation we have neglected
the time derivatives and the pressure terms.

The linearization of the metric field equations is a bit more
complicated because we need to establish suitable gauge
conditions. Since the background is Minkowskian, the perturbed
Ricci tensor is given by
\begin{equation}
\delta
R_{\mu\nu}\equiv\frac{1}{2}\left(\partial_\mu\partial_\lambda
\tilde{h}^\lambda_\nu+\partial_\nu\partial_\lambda
\tilde{h}^\lambda_\mu\right)-\frac{1}{2}\vec{\nabla}^2h_{\mu\nu} \,,
\end{equation}
where $\tilde{h}^\lambda_\nu\equiv
{h}^\lambda_\nu-(1/2)\delta^\lambda_\nu h^\alpha_\alpha$. The term
$\nabla_\mu\nabla_\nu\phi$ on the righthand side of
Eq.~(\ref{eq:evol-gab}) can be combined with the terms
$\partial_\mu\partial_\lambda \tilde{h}^\lambda_\nu$ to give the
following gauge conditions
\begin{equation}
\partial_\lambda \tilde{h}^\lambda_\mu-\frac{1}{1+\phi_0}\partial_\mu \varphi=0\,.
\end{equation}
With this choice, the linearized equations for the metric become
\begin{equation}\label{eq:linear-gab}
-\frac{1}{2}\vec{\nabla}^2h_{\mu\nu}=\frac{1}{1+\phi_0}\left(T_{\mu\nu}-\frac{1}{2}T
\eta_{\mu\nu}\right)+\frac{V_0+\vec{\nabla}^2\varphi}{2(1+\phi_0)}\eta_{\mu\nu}\,.
\end{equation}
For consistency, to this order $T_{00}=\rho$, $T_{ij}=0$,
$T=-\rho$, which leads to
\begin{eqnarray}\label{eq:linear-h00}
\vec{\nabla}^2\left(h_{00}^{(2)}-\frac{\varphi^{(2)}}{1+\phi_0}\right)&=& -\frac{1}{1+\phi_0}\left(\rho-V_0\right), \\
\vec{\nabla}^2\left(h_{ij}^{(2)}+\frac{\varphi^{(2)}}{1+\phi_0}\delta_{ij}\right)&=&
-\frac{\delta_{ij}}{1+\phi_0}\left(\rho+V_0\right)\,.
\label{eq:linear-hij}
\end{eqnarray}

Before solving Eqs.~(\ref{eq:linear-phi}), (\ref{eq:linear-h00}),
and (\ref{eq:linear-hij}), it is worth noting that while the
connection equation (\ref{eq:var-phi}) is invariant under constant
rescalings of the field $\phi\to c \phi$, the other field
equations do not share this invariance. This is manifest in the
combinations $(1+\phi_0)$ in the above perturbation equations.


Using the generic solution
\begin{equation}
(\vec{\nabla}^2-m^2)f=-\rho  \ \Rightarrow \ f=\frac{1}{4\pi}\int
{\rm d}^3\vec{x}'\frac{\rho(t,\vec{x}')}{|\vec{x}-\vec{x}'|}e^{-m|\vec{x}-\vec{x}'|
} ,
\end{equation}
we find that
\begin{eqnarray}
\varphi^{(2)}(t,\vec{x})&=&\frac{\kappa^2\phi_0}{4\pi}\int {\rm d}^3\vec{x}'\frac{\rho(t,\vec{x}')}{3|\vec{x}-\vec{x}'|}e^{-m_\varphi|\vec{x}-\vec{x}'| }, \\
h_{00}^{(2)}(t,\vec{x})&=&\frac{\kappa^2}{4\pi(1+\phi_0)}\int
{\rm d}^3\vec{x}'\frac{\rho(t,\vec{x}')}{|\vec{x}-\vec{x}'|}
\left(1+\frac{\phi_0}{3}e^{-m_\varphi|\vec{x}-\vec{x}'| }\right) +\frac{V_0}{(1+\phi_0)}\frac{|\vec{x}-\vec{x}_2|^2}{6},\\
h_{ij}^{(2)}(t,\vec{x})&=&\Bigg[\frac{\kappa^2}{4\pi(1+\phi_0)}\int
{\rm d}^3\vec{x}'\frac{\rho(t,\vec{x}')}{|\vec{x}-\vec{x}'|}
\left(1-\frac{\phi_0}{3}e^{-m_\varphi|\vec{x}-\vec{x}'| }\right)
-\frac{V_0}{(1+\phi_0)}\frac{|\vec{x}-\vec{x}_2|^2}{6}\Bigg]\delta_{ij}.
\end{eqnarray}

In spherical symmetry and far from the sources, the above
equations become
\begin{eqnarray}
\varphi(r)&=&\frac{\kappa^2\phi_0}{12\pi}\frac{M}{r}e^{-m_\varphi r},\label{cor1} \\
h_{00}^{(2)}(r)&=& \frac{2G_{\rm eff} M}{r} +\frac{V_0}{(1+\phi_0)}\frac{r^2}{6},\label{cor2}\\
h_{ij}^{(2)}(r)&=& \left( \frac{2\gamma G_{\rm eff} M}{r}
-\frac{V_0}{(1+\phi_0)}\frac{r^2}{6} \right) \delta_{ij} \label{cor3},
\end{eqnarray}
where we have denoted
\begin{eqnarray}
G_{\rm eff}&\equiv & \frac{\kappa^2}{8\pi (1+\phi_0)}\left(1+\frac{\phi_0}{3}e^{-m_\varphi r}\right), \\
\gamma &\equiv & \frac{\left[1+\phi_0\exp \left(-m_{\varphi} r\right)/3\right]}{\left[1-\phi_ 0\exp \left(-m_{\varphi } r\right)/3\right]},\label{gamma0} \\
m_\varphi^2 &\equiv
&\left.\frac{2V-V_{\phi}-\phi(1+\phi)V_{\phi\phi}}{3}\right|_{\phi=\phi_0}\,. \label{mass}
\end{eqnarray}
These results represent the standard post-Newtonian metric up to
second order for this class of theories.

We emphasize a striking feature of $f(X)$ gravity. Note that in
$f(R)$ gravity, to obtain $\gamma\approx 1$ from (\ref{gamma0}),
there is only one possibility \cite{Olmo:2005zr,Olmo:2005hc}, namely,
$m_\varphi r \gg 1$ from millimetres to astronomical scales, i.e.,
the range of the scalar interaction, $1/m_\varphi$, should be smaller than a few millimetres.
In the current case, however, there are two possibilities to
obtain $\gamma\approx 1$. The first one is the same as in $f(R)$
theories and involves a very massive scalar field. The second
possibility implies a small value $\phi_0$. If $\phi_0\ll 1$, then
the Yukawa-type corrections are very small regardless of the
magnitude of $m_\varphi$. This could allow for the existence of a
long-range scalar field able to modify the cosmological and
galactic dynamics, but leaving unaffected the Solar System. Subtle modifications
could in the most optimistic case be detected as anomalies in the local gravitational field \cite{bah2}.

\subsection{Galactic phenomenology: Stable circular orbits of test particles around galaxies}


The most direct method for studying the gravitational field inside a spiral galaxy is provided by the galactic rotation
curves. They are obtained by measuring the frequency shifts $z$ of the 21-cm radiation emission from the neutral
hydrogen gas clouds. The 21-cm radiation also originates from stars. The 21-cm background from the epoch of reionization is a promising cosmological probe: line-of-sight velocity fluctuations distort redshift, so brightness fluctuations in Fourier space depend upon angle, which linear theory shows can separate cosmological from astrophysical information (for a recent review see \cite{21cm}). Instead of using $z$ the resulting redshift is presented by astronomers in terms of a velocity
field $v_{tg}$ \cite{dm,dm2}. 

In the following, we will assume that the gas clouds behave like test particles, moving in the static and
spherically symmetric geometry around the galaxy. Without a significant loss of generality, we assume that the gas
clouds move in the galactic plane $\theta =\pi /2$, so that their four-velocity is given by $u^{\mu }=\left( \dot{t},
\dot{r},0,\dot{\phi}\right)$, where the overdot stands for derivation with respect to the affine parameter $s$. In this subsection
we find it illustrative to restore the units of $c$.

The static spherically symmetric metric outside the galactic baryonic mass distribution is given by the following line element 
\begin{equation} 
{\rm d}s^{2}=-e^{\nu \left( r\right) }c^2{\rm d}t^{2}+e^{\lambda \left( r\right)
}{\rm d}r^{2} + r^{2}\left( {\rm d}\theta ^{2}+\sin ^{2}\theta {\rm d}\phi ^{2}\right) ,
\label{line}
\end{equation}
where the metric coefficients $\nu (r)$ and $\lambda (r)$ are functions of the radial coordinate $r$ only. The motion of a
test particle in the gravitational field with the metric given by Eq.~(\ref{line}), is described by the Lagrangian \cite{Nuc01,Nuc01b}
\begin{equation}
L=\left[ e^{\nu \left( r\right) }\left( \frac{cdt}{ds}\right)
^{2}-e^{\lambda \left( r\right) }\left( \frac{dr}{ds}\right)
^{2}-r^{2}\left( \frac{d\Omega }{ds}\right) ^{2}\right] ,
\end{equation}
where $d\Omega ^{2}=d\theta ^{2}+\sin ^{2}\theta d\phi ^{2}$, which simplifies to $d\Omega ^{2}=d\phi ^{2}$
along the galactic plane $\theta=\pi /2$. From the Lagrange equations it follows that we have two constants of
motion, namely, the energy $E$ per unit mass, and the angular momentum $l$ per unit mass, given by $E=e^{\nu (r)}c^3\dot{t}$ and
$l=cr^{2}\dot{\phi}$, respectively. The normalization condition for the four-velocity $u^{\mu }u_{\mu }=-1$ gives
$1=e^{\nu \left( r\right) }c^2\dot{t}^{2}-e^{\lambda (r)}\dot{r}^{2}-r^{2}\dot{\phi}^{2}$, from which, with the
use of the constants of motion, we obtain the energy of the particle as
\begin{equation}
\frac{E^{2}}{c^2}=e^{\nu +\lambda }\dot{r}^{2}+e^{\nu }\left( \frac{l^{2}}{c^2r^{2}}%
+1\right) .  \label{energy}
\end{equation}

From Eq.~(\ref{energy}) it follows that the radial motion of the test particles is analogous to that of particles in
Newtonian mechanics, having a  velocity $\dot{r}$, a position dependent effective mass $m_{\rm eff}=2e^{\nu +\lambda }$, and an energy $E$. In addition to this, the test particles move in an effective potential provided by the following relationship
\begin{equation}
V_{\rm eff}\left( r\right) =e^{\nu (r)}\left( \frac{l^{2}}{c^2r^{2}}+1\right) .
\end{equation}
The conditions for circular orbits, namely, $\partial V_{\rm eff}/\partial r=0$ and $\dot{r}=0$ lead to
\begin{equation}\label{cons}
l^{2}=\frac{c^2}{2}\frac{r^{3} \nu^{\prime } }{1-r\nu ^{\prime }/2}\,, \quad
\frac{E^{2}}{c^4}=\frac{e^{\nu } }{1-r\nu ^{\prime }/2}\,,
\end{equation}
respectively.
Note that the spatial three-dimensional  velocity is given by
\begin{equation}
v^{2}(r)=e^{-\nu }\left[ e^{\lambda }\left( \frac{{\rm d}r}{{\rm d}t}\right)
^{2}+r^{2}\left( \frac{{\rm d}\Omega }{{\rm d}t}\right) ^{2}\right] .
\end{equation}
For a stable circular orbit $dr/dt=0$, and the tangential velocity of the
test particle can be expressed as
\begin{equation}
v_{tg}^{2}(r)=e^{-\nu }r^{2}\left( \frac{{\rm d}\Omega }{{\rm d}t}\right) ^{2}=e^{-\nu }r^{2}\left( \frac{{\rm d}\Omega }{{\rm d}s}\right) ^{2}\left(\frac{{\rm d}s}{{\rm d}t}^2\right).
\end{equation}
In terms of the conserved quantities, and along the galactic plane $\theta =\pi /2$, the angular velocity is given by
\begin{equation}
\frac{v_{tg}^{2}(r)}{c^2}=c^2\frac{e^{\nu }}{r^{2}}\frac{l^{2}}{E^{2}}\,,
\end{equation}
and taking into account Eq.~(\ref{cons}), we finally obtain the following relationship \cite{Nuc01,Nuc01b}
\begin{equation}
\frac{v_{tg}^{2}(r)}{c^2}=\frac{r \nu ^{\prime }}{2}\,.  \label{vtg}
\end{equation}
Therefore, once the tangential velocity of test particles is known, the metric function $\nu(r)$ outside the galaxy can
be obtained as
\be
\nu (r)=2\int{\frac{v_{tg}^2(r)}{c^2}\frac{{\rm d}r}{r}}\,.
  \label{metricnu}
\ee
The tangential velocity $v_{tg}/c$ of gas clouds moving like test particles around the center of a galaxy is not directly
measurable, but can be inferred from the redshift $z_{\infty }$ observed at spatial infinity, for which
$1+z_{\infty}=\exp \left[ \left( \nu _{\infty }-\nu \right) /2 \right] \left( 1\pm v_{tg}/c\right) /\sqrt{1-v_{tg}^{2}/c^2}$ \cite{Nuc01,Nuc01b}.
Due to the non-relativistic velocities of the gas clouds, with $v_{tg}/c\leq \left( 4/3\right) \times 10^{-3}$, we observe
that $v_{tg}/c\approx z_{\infty }$, as the first part of a geometric series. The observations show that at distances large
enough from the galactic center the tangential velocities assume a constant value, i.e., $v_{tg}/c\approx $ constant \cite{dm,dm2}. In the latter regions of the constant tangential velocities, Eq. (\ref{metricnu}) can be readily integrated to provide the following metric tensor component
\be\label{nu}
e^{\nu }=\left(\frac{r}{R_{\nu }}\right)^{2v_{tg}^2/c^2}\approx 1+2\frac{v_{tg}^2}{c^2}\ln\left(\frac{r}{R_{\nu }}\right),
\ee
where $R_{\nu }$ is an arbitrary constant of integration. If we match the metric given by Eq.~(\ref{nu}) with the Schwarzschild metric on the surface of the galactic baryonic matter distribution, having a radius $R_B$, $\left.e^{\nu}\right|_{r=R_B}=1-2GM_B/c^2R_B$, we obtain the following relationship
\be
R_{\nu}=\frac{R_B}{\left(1-2GM_B/c^2R_B\right)^{c^2/2v_{tg}^2}}.
\ee

An important physical requirement for the circular orbits of the test particle around galaxies is that they
must be stable. Let $r_{0}$ be the radius of a circular orbit and consider a perturbation of it of the form
$r=r_{0}+\delta $, where $\delta \ll r_{0}$ \citep{La03}. Taking expansions of $V_{\rm eff}\left(
r\right) $ and $e^{\nu +\lambda}$  about $r=r_{0} $, it follows from Eq.~(\ref{energy})
that
\begin{equation}
\ddot{\delta}+\frac{1}{2} e^{\nu \left( r_{0}\right)
+\lambda \left( r_{0}\right) }V_{\rm eff}^{\prime \prime }\left( r_{0}\right)
\delta =0.
\end{equation}
The condition for stability of the simple circular orbits requires $V_{\rm eff}^{\prime \prime }\left(
r_{0}\right) >0$ \citep{La03}. Hence, with the use of the condition $V_{\rm eff}^{\prime}\left (r_0\right)=0$, we obtain the condition of the stability of the orbits as $\left[3\nu '+r\nu ''>r\nu '^2/2\right]|_{r=r_0}$. By taking into account Eq.~(\ref{vtg}), it immediately follows stable circular orbits always exist for massive test particles.

\subsubsection{Galactic geometry and tangential velocity curves in hybrid metric-Palatini gravity}\label{sect4}

The rotation curves only determine one, namely $\nu (r)$, of the two unknown metric functions, $\nu (r)$ and $\lambda (r)$, which are required to describe the gravitational field of the galaxy \cite{Capozziello:2013yha}. Hence, in order to determine $\lambda(r)$ we proceed to solve the gravitational field equations for the hybrid metric-Palatini gravitational theory outside the baryonic matter distribution. This allows us to take all the components of the ordinary matter stress-energy tensor as being zero. Furthermore, our task becomes easier when we restrict to perturbative weak-field treatment.


The weak field limit of the gravitational theories at the Solar System level is usually obtained by using isotropic coordinates as above in Section \ref{weakfield}. However, it is useful to apply Schwarzschild coordinates in studying exact solutions and in the context of galactic dynamics, and this is what we do here.
.
We assume that the gravitational field inside the halo is weak, so that $\nu (r)\sim \lambda (r)\sim (v_{tg}/c)^2$,
which allows us to linearise the gravitational field equations retaining only terms linear in $(v_{tg}/c)^2$ and again consider the scalar field as $\phi
=\phi _0+\varphi $, where $\varphi \ll 1$ is a small perturbation around the background value $\phi _0>0$.
The Klein-Gordon equation was already solved in Section \ref{weakfield} at this limit and the result was that the interaction range is given
by $r_\varphi=1/m_\varphi$, where $m_\varphi$ is given in Eq. (\ref{mass}). In a spherically symmetric configuration, the general solution then has the form
\be
\varphi (r)=\Psi _0\frac{e^{-r/r_{\varphi}}}{r}\,,
\ee
where $\Psi _0$ is an integration constant. Comparing this expression with the results obtained in \cite{Harko:2011nh} for the weak-field limit (taking into account the transformation from isotropic to Schwarzschild coordinates), we find that
\be\label{Psi0}
\Psi_0=-\frac{2GM_B}{c^2}\phi_0\frac{e^{R_B/ r_\varphi }}{3}<0\,,
\ee
where $M_B$ and $R_B$ are the mass and the radius of the galactic baryonic distribution, respectively.

Within this linear approximation the stress-energy tensor of the scalar field is given by
\be
T_{\mu \nu }^{(\phi)}=\frac{1}{\kappa ^2}\left[\nabla _{\mu }\nabla _{\nu }\varphi +\left(\alpha \varphi +\beta \right)g_{\mu \nu}\right]\,,
\ee
where $\alpha$ and $\beta$ are defined by
\be
\alpha =\frac{1}{r_{\varphi }^2}- \frac{1}{2}V'\left(\phi _0\right)\ , \qquad \beta =-\frac{1}{2}V\left(\phi _0\right)\,.
\ee
Therefore the linearized gravitational field equations take the form
\bea\label{f1}
\frac{1}{r^2}\frac{d}{dr}\left(r\lambda\right)=\alpha \varphi +\beta\,,
  \\
\label{f2}
-\frac{\nu '}{r}+\frac{\lambda }{r^2}=\varphi ''+\alpha \varphi +\beta\,,
   \\
\label{f3}
-\frac{1}{2}\left(\nu ''+\frac{\nu '-\lambda '}{r}\right)=\alpha \varphi +\beta\,.
\eea
 Eq.~(\ref{f1}) can be immediately integrated to provide
\be
\lambda (r)=\frac{\lambda_0}{r}+\frac{1}{r}\int ^r{\left(\alpha \varphi +\beta \right)\tilde{r}^2{\rm d}\tilde{r}}
=\frac{\lambda_0}{r}+\frac{\beta}{3}r^2-\frac{\alpha r_\varphi^2 \Psi_0 e^{-r/r_\varphi}}{r}\left(1+\frac{r}{r_\varphi}\right) \,,
\ee
where $\lambda_0$ is an integration constant. Comparing again with the results obtained in \cite{Harko:2011nh} for the weak-field limit, we find that $\lambda_0=2GM/c^2$.
The tangential velocity of the test particles in stable circular orbits moving in the galactic halo can be derived immediately from Eq.~(\ref{f2}), and is given by
\be
\frac{v_{tg}^2}{c^2}=\frac{r\nu '}{2}=\frac{\lambda }{2}-r^2\frac{\varphi ''}{2}-\frac{\alpha }{2}r^2\varphi-\frac{\beta}{2}r^2,
\ee
which in terms of the solutions found above becomes
\bea
\frac{v_{tg}^2}{c^2}&=&\frac{V_0}{6} r^2 + \frac{GM_B}{c^2r}-\frac{\Psi_0 e^{-r/r_\varphi}}{2r}
\left[\left(1+\frac{r}{r_\varphi}\right)(2+\alpha r_\varphi^2)+\frac{r^2}{r_\varphi^2}(1+\alpha r_\varphi^2)\right],
\eea
where $V_0=-\beta =V\left(\phi _0\right)/2$.
The term proportional to $r^2$ corresponds to the cosmological background, namely the de Sitter geometry, and we assume that it has a negligible contribution on the tangential velocity of the test particles at the galactic level.

On the surface of the baryonic matter distribution the tangential velocity must satisfy the boundary condition
\be
\frac{v_{tg}^2\left(R_B\right)}{c^2}\approx \frac{GM_B}{c^2R_B}\,,
\ee
 which, with the use of Eq.~(\ref{Psi0}), gives the following constraint on the parameters of the model,
\be
\left(1+\frac{R_B}{r_\varphi}\right)(2+\alpha r_\varphi^2)+\frac{R_B^2}{r_\varphi^2}(1+\alpha r_\varphi^2)\approx 0\,.
\ee
In order to satisfy the above condition would require that $-2<\alpha r_{\varphi }^2<-1$, or, equivalently,
\be
V'\left(\phi _0\right)>0\,, \quad
2<\frac{1}{2}V'\left(\phi _0\right)r_{\varphi }^2<3\,.
\ee

In the regions near the galactic baryonic matter distribution, where $R_B\leq r \ll r_{\varphi}$, we have $e^{-r/r_{\varphi}}\approx 1$, to a very good approximation. Hence in this region the tangential velocity can be approximated as
\bea
\frac{v_{tg}^2}{c^2}&\approx&\frac{2GM_B-c^2\Psi_0\left(\alpha r_{\varphi}^2+2\right)}{2c^2r}-\Psi_0\frac{\alpha r_{\varphi }^2+2}{2r_{\varphi }}
-\frac{\Psi _0}{2r_{\varphi }^2}\left(1+\alpha r_{\varphi }^2\right)r, \qquad R_B\leq r \ll r_{\varphi}\,.
\eea
If the parameters of the model satisfy the condition
\bea
2GM_B-c^2\Psi_0\left(\alpha r_{\varphi}^2+2\right)\approx 0\,,
\eea
the term proportional to $1/r$ becomes negligible, while for small values of $\Psi _0$, and $\left|\alpha r_{\varphi ^2}\right|\approx 1$, the term
proportional to $r$ can also be neglected. Therefore for the tangential velocity of test particles rotating in the
galactic halo we obtain
\be
\frac{v_{tg}^2}{c^2}\approx -\Psi_0\frac{\alpha r_{\varphi }^2+2}{2r_{\varphi}}\approx -\frac{\Psi _0\alpha
r_{\varphi}}{2},  \qquad R_B\leq r \ll r_{\varphi}.
\ee
Since according to our assumptions, $r_{\varphi} \gg 1$, the coefficient $\alpha $ can be approximated as $\alpha
\approx -V'\left(\phi _0\right)/2$, which provides for the rotation curve, in the constant velocity region, the
following expression
\be
\frac{v_{tg}^2}{c^2}\approx \frac{\Psi _0V'\left(\phi _0\right) r_{\varphi}}{4}, \qquad R_B\leq r  \ll r_{\varphi}.
\ee
Since $\Psi_0<0$, the scalar field potential must satisfy the condition $V'\left(\phi _0\right)<0$. In the first order of approximation, with $\exp\left(-r/r_{\varphi}\right)\approx1-r/r_{\varphi}$, for the tangential velocity we obtain the expression
\be
\frac{v_{tg}^2}{c^2}\approx \frac{2GM_B-c^2\Psi _0\left(\alpha r_{\varphi}^2+2\right)}{2c^2r}+\frac{\Psi _0}{2r_{\varphi }^2}r+\frac{\Psi_0\left(\alpha r_{\varphi }^2+1\right)}{2r_{\varphi }^2}r^2.
\ee
Alternatively, in general we can write the tangential velocity  as follows,
\bea\label{vfin}
\frac{v_{tg}^2}{c^2}&=&\frac{V_0}{6} r^2 + \frac{GM_B}{c^2r}\Bigg \{1+\frac{2\phi_0}{3}e^{\frac{GM_B/c^2-r}{r_\varphi}}\Bigg [\left(1+\frac{r}{r_\varphi}\right)
(2+\alpha r_\varphi^2)+\frac{r^2}{r_\varphi^2}(1+\alpha r_\varphi^2)\Bigg ]\Bigg\}.
\eea
As compared to our previous results, in this representation we have $e^{\frac{GM_B/c^2-r}{r_\varphi}}$ instead of $e^{\frac{R_B-r}{r_\varphi}}$. Since we are working in a regime in which $R_B\ll r_\varphi$, the choice of the constants $R_B$ or $M_B$ does not seem very  relevant, since it just amounts to a rescaling of $\phi_0$. From now on we will also assume that $e^{GM_B/c^2r_\varphi}\approx 1$.

From the above equation we want to find the constraints on the model parameters that arise from the expected behavior at different scales. For that purpose, it is convenient to write the equation, equivalently, as follows:
\bea
\frac{v_{tg}^2}{c^2}&=& \frac{GM_B}{c^2r}\left[1+\frac{2\phi_0}{3}(2+\alpha r_\varphi^2)e^{-\frac{r}{r_\varphi}}\right] \nonumber\\
&& + \frac{GM_B}{c^2r_\varphi} \left(2+\alpha r_\varphi^2\right) e^{-\frac{r}{r_\varphi}}+\frac{GM_B}{c^2r_\varphi} (1+\alpha r_\varphi^2) \left(\frac{r}{r_\varphi}\right) e^{-\frac{r}{r_\varphi}} .
\eea
At intermediate scales, the asymptotic tangential velocity tends to a constant.
If we expand the exponential as $e^{-\frac{r}{r_\varphi}}\approx 1-r/r_\varphi$, hen we obtain the following three constraints on the free parameters of the model,
\bea
&&a) \;\;1+\frac{2\phi_0}{3}\left(2+\alpha r_\varphi^2 \right)\approx 0, \\
&&b) \;\;  \left(2+\alpha r_\varphi^2 \right)\left(1-\frac{2\phi_0}{3}\right)\approx C= {\rm constant},\\
&&c) \; \;\frac{GM_B}{c^2r_\varphi}\left(\frac{r}{r_\varphi}\right) \ll |C|.
\eea
With increasing $r$, and by assuming that the condition $r \ll r_{\varphi}$ still holds, the rotation curves will decay, at very large distances from the galactic center, to the zero value.

\subsubsection{On astrophysical tests of hybrid metric-Palatini gravity at the galactic level}\label{sect5}

In \cite{Capozziello:2013yha}, some observational astrophysical tests of hybrid metric-Palatini gravity at the galactic level were discussed. 
More specifically, a comparison of the theoretical predictions of the model with a sample of rotation curves of low surface brightness galaxies was analysed. Indeed, the predictions of the theoretical model with the observational results show that the contribution of the scalar field energy density to the tangential velocity of the test particles can explain the existence of a constant rotational velocity region around the baryonic matter, without requiring the presence of the dark matter. The possibility of observationally determining the functional form of the scalar field $\varphi $ by using the velocity dispersion of stars in galaxies, and the red and blue shifts of gas clouds moving in the galactic halo could also be worth considering. It can be shown that one can constrain the explicit functional form of the scalar field, and the free parameters of the model, in order to adequately fit the observational data. We refer the reader to \cite{Capozziello:2013yha} for more details. Of course, was particle matter detected directly, we could exclude the gravitational $f(X)$ alternative to dark matter.

\subsection{Galactic clusters: The generalized virial theorem in hybrid metric-Palatini gravity}\label{sec:c}

A large number of astronomical and astrophysical observations confirm that galaxies form complex hierarchical structures, in which galaxies concentrate in large objects, called clusters of galaxies, bounded by the gravitational interaction. The total mass of the galaxy cluster ranges from $10^{13} M_{\odot }$ for the so-called groups (formed by a few hundred of galaxies) up to a few $10^{15} M_{\odot}$ for very large clusters, containing thousands of galaxies. From a morphological point of view galactic clusters are usually by a main component, which is regular and centrally peaked~\cite{ReBo02,Ar05}. For fundamental physics research the main importance of the galactic clusters consists in the fact that they are considered to be ``dark matter'' dominated astrophysical objects. Moreover, their formation and evolution is almost entirely controlled  by the gravitational force, a property which allows the testing of different dark matter models. On the other hand the mass distribution inside the clusters is fully determined by the initial conditions of the mass distribution that originate in the early universe~\cite{Sch01,Sch01b,Sch01c}, thus allowing the use of galaxy cluster properties to also test cosmological models. In this subsection we discuss the possibility of observationally testing the metric-Palatini gravity theory by using observational evidence from galaxy clusters \cite{Capozziello:2012qt}.

\subsubsection{Galaxy cluster as a system of identical and collisionless point particles}

As a first step necessary to obtain our main result, consisting in the generalization of the relativistic virial theorem for galaxy clusters in the hybrid metric-Palatini gravitational we write down the gravitational field equations and the Boltzmann for a static and spherically symmetric distribution of matter. We adopt a simplified physical model, in which the galactic cluster consists of a self-gravitating system of identical, collisionless point particles (the galaxies) in random dynamical motion. The metric will thus be described by (\ref{line}), and the fluid by a distribution function $f_B$ that obeys the general relativistic Boltzmann equation  \cite{HaCh07}. 

The energy-momentum tensor of the matter in the cluster, is thus determined by the distribution function $f_B$, and its components are given by the equation~\cite{Li66,Li66b}
\be \label{source}
T_{\mu \nu }=\int f_B\, m\, u_{\mu }u_{\nu }\;{\rm d}u,
\ee
where $m$ is the mass of the particle (galaxy), $u_{\mu }=\left(u_t,u_r,u_{\theta},u_{\varphi}\right)$ is the
corresponding galactic four-velocity, with $u_t$ denoting the temporal component. 
Finally, by ${\rm d}u ={\rm d}u_{r}{\rm d}u_{\theta }{\rm d}u_{\varphi }/u_{t}$ we denote the invariant volume element in the velocity space. Alternatively, the
energy-momentum tensor $T_{\mu \nu }$ describing the matter distribution in a cluster of galaxies can be
represented in terms of an effective energy density $\rho_{\mathrm{eff}}$, and of two
effective anisotropic thermodynamic pressures,  the radial $p_{\mathrm{eff}}^{(r)}$ and the
tangential $p_{\mathrm{eff}}^{(\perp)}$ pressure, respectively.  These thermodynamic parameters of the cluster are obtained by averaging over the matter and velocity distributions, and are given by
\be\label{164}
\rho_{\mathrm{eff}} = \rho \left\langle u_{t}^{2}\right\rangle, p_{%
\mathrm{eff}}^{(r)}=\rho \left\langle u_{r}^{2}\right\rangle,
p_{\mathrm{eff}}^{(\perp)} = \rho \left\langle u_{\theta
}^{2}\right\rangle= \rho \left\langle u_{\varphi }^{2}\right\rangle,
\ee
where $\rho $ is the mass density of the ordinary baryonic matter, and $\left\langle u_{i}^{2}\right\rangle $, $i=t,r,\theta ,\varphi $ denotes the average
value of $u_{i}^{2}$, $i=t,r,\theta ,\varphi $, representing the square of the components of the four-velocities of the galaxies in the cluster~\cite{Ja70}.
The full field equations for the metric (\ref{line}) with the source (\ref{164}) were listed in Ref. \cite{Capozziello:2012qt}, but here we shall only need their trace:
\begin{multline}
e^{-\lambda }\left( \frac{\nu ^{\prime \prime }}{2}+\frac{\nu ^{\prime 2}}{4}
+\frac{\nu ^{\prime }}{r}-\frac{\nu ^{\prime }\lambda ^{\prime }}{4}\right)
=4\pi \frac{G}{1+\phi }\rho \left\langle u^{2}\right\rangle \\
+\frac{1}{1+\phi } V\left( \phi \right) +\frac{1}{1+\phi }\left( 2\nabla
_{t}\nabla ^{t}+ \nabla_\alpha\nabla^\alpha \right) \phi -\frac{3}{\phi (1+\phi )}\nabla _t\phi \nabla ^t
\phi,  \label{ff}
\end{multline}
where we have denoted $\langle u^{2}\rangle =\langle u_{t}^{2}\rangle +\langle
u_{r}^{2}\rangle +\langle u_{\theta }^{2}\rangle +\langle u_{\varphi
}^{2}\rangle$.

Since in the following we are interested only in astrophysical applications at the extra-galactic cluster
scale, we will adopt a Newtonian type approximation, which consists in assuming that the deviations from standard general relativity
(corresponding to the background value $\phi =0$) are small for the systems we are considering. This approximation implies that $\phi \ll 1$. Thus,  Eq.~(\ref{ff}) can be approximated as
\be
e^{-\lambda }\left( \frac{\nu ^{\prime \prime }}{2}+\frac{\nu ^{\prime 2}}{4}
+\frac{\nu ^{\prime }}{r}-\frac{\nu ^{\prime }\lambda ^{\prime }}{4}\right)
\simeq 4\pi G\rho \left\langle u^{2}\right\rangle +4\pi G\rho _{\phi }^{(eff)},
\label{ff1}
\ee
where
\be\label{approx}
4\pi G\rho _{\phi }^{(eff)} \simeq  V\left( \phi \right) +\left( 2\nabla
_{t}\nabla ^{t}+ \nabla_\alpha\nabla^\alpha \right) \phi -\frac{3}{\phi }\nabla _t\phi \nabla ^t\phi ,
\ee
corresponds to an effective, geometric type  "energy" of the scalar field in the hybrid metric-Palatini gravitational model.

\subsubsection{The relativistic Boltzmann equation}

Next we proceed to the second step in the derivation of the virial theorem for galaxy clusters and determine the general relativistic Boltzmann equation that governs the evolution of the galactic distribution function $f_B$.

A basic result in statistical physics is the transport equation for the distribution function  for a system of particles in a
curved arbitrary Riemannian space-time. This transport equation is given by the Boltzmann equation without collision term, and which can be formulated as ~\cite{Li66,Li66b}
\be  \label{distr}
\left( p^{\alpha }\frac{\partial }{\partial x^{\alpha }}-p^{\alpha }p^{\beta
}\Gamma _{\alpha \beta }^{i}\frac{\partial }{\partial p^{i}}\right) f_B=0\,,
\ee
where $p^{\alpha }$ is the four-momentum of the galaxy (particle), and $\Gamma _{\alpha \beta }^{i}$ are the Christoffel symbols associated to the metric. An important consequence of the collissionless Boltzmann equation is that the local phase space density, as measured by an observer in a frame co-moving with a galaxy, is conserved.

An important simplification of the mathematical formulation of the Boltzmann equation can be achieved by introducing an appropriately chosen
orthonormal frame, or tetrad $\e_{\mu }^{a}(x)$, $a=0,1,2,3$. The tetrad fields vary
smoothly over some coordinates neighborhood $U$, and they satisfy the general condition $%
g^{\mu\nu} \e_{\mu}^{a} \e_{\nu}^{b} =\eta^{ab}$ for all $x\in U$, where $\eta ^{ab}$ denotes the Minkowski metric tensor~\cite
{Li66,Li66b,Ja70}. A basic property of the tetrad fields is that any tangent vector $p^{\mu}$ defined at an arbitrary point $x$ can be represented as $%
p^{\mu}=p^{a}\e_{a}^{\mu}$, a relation which defines the tetrad components $p^{a}$.

For the spherically symmetric line element given by Eq.~(\ref{line}), the frame of orthonormal vectors can be chosen in an appropriate way as~\cite
{Li66,Li66b,Ja70}:
\be
\e_{\mu}^{0} = e^{\nu /2}\delta _{\mu}^{0}, \quad \e_{\mu}^{1} =
\e^{\lambda /2}\delta _{\mu }^{1}\,, \quad
\e_{\mu}^{2} = r\delta_{\mu}^{2}\,, \quad  \e_{\mu}^{3} = r\sin \theta
\delta_{\mu}^{3}\,.
\ee
The tetrad components of the four velocity are $u^{a}=u^{\mu}\e_{\mu }^{a}$. In the tetrad components corresponding to our present choices the relativistic Boltzmann equation Eq.~(\ref{distr}) is given by
\begin{equation}
u^{a}\e_{a}^{\mu}\frac{\partial f_B}{\partial x^{\mu}}+
\gamma_{bc}^{a}u^{b}u^{c}\frac{\partial f_B}{\partial u^{a}}=0,  \label{tetr}
\end{equation}
where the distribution function $f_B=f_B(x^{\mu },u^{a})$ is a function of $x^{\mu}$ and $u^a$, respectively, and $\gamma
_{bc}^{a}=\e_{\mu ;\nu }^{a}\e_{b}^{\mu}\e_{c}^{\nu}$ are the Ricci rotation
coefficients~\cite{Li66,Li66b,Ja70}. Due to the spherical symmetry of the problem the distribution function depends only on the radial coordinate $r$,
and hence Eq.~(\ref{tetr}) becomes~\cite{Ja70}
\begin{eqnarray}
u_{1}\frac{\partial f_B}{\partial r}- \left(\frac{1}{2}u_{0}^{2}\frac{%
\partial \nu}{\partial r}- \frac{u_{2}^{2}+u_{3}^{2}}{r}\right) \frac{%
\partial f_B}{\partial u_{1}}
-\frac{1}{r}u_{1}\left( u_{2}\frac{\partial f_B}{\partial u_{2}} +u_{3}\frac{%
\partial f_B}{\partial u_{3}}\right)
    \nonumber \\
-\frac{1}{r}e^{\lambda /2}u_{3}\cot \theta \left( u_{2} \frac{\partial f_B}{%
\partial u_{3}}-u_{3} \frac{\partial f_B}{\partial u_{2}}\right) = 0.
\label{tetr1}
\end{eqnarray}


 Due to the spherical symmetry of our astrophysical system  the coefficient of $\cot \theta $ in Eq.~(\ref{tetr1}) must be zero. From a mathematical point of view this implies that the distribution function $f_B$ is only a function of $r$, $u_{1}$ and $%
u_{2}^{2}+u_{3}^{2}$. As a next steps in our analysis we multiply Eq.~(\ref{tetr1}) by $m u_{r} {\rm d}u $,
and we integrate over the velocity space. Then,  by taking into account that the distribution function $f_B$
vanishes sufficiently rapidly as the velocities tend to $\pm \infty $, we
find the equation
\bea
&&r\frac{\partial}{\partial r}\left[\rho\left\langle u_{1}^{2}\right\rangle%
\right]+ \frac{1}{2}\rho \left[ \left\langle u_{0}^{2}\right\rangle +
\left\langle u_{1}^{2}\right\rangle\right] r\frac{\partial \nu }{\partial r}
-\rho \left[ \left\langle u_{2}^{2}\right\rangle +\left\langle
u_{3}^{2}\right\rangle -2\left\langle u_{1}^{2}\right\rangle \right] =0.
\label{tetr2}
\eea
Now we multiply Eq.~(\ref{tetr2}) by $4\pi r^{2}$, and by integrating over the entire cluster volume, corresponding to a radius $R$, gives~\cite
{Ja70}
\bea
&&\int_{0}^{R}4\pi \rho \left[ \left\langle u_{1}^{2}\right\rangle
+\left\langle u_{2}^{2}\right\rangle +\left\langle u_{3}^{2}\right\rangle%
\right] r^{2}{\rm d}r
-\frac{1}{2}\int_{0}^{R}4\pi r^{3}\rho \left[ \left\langle
u_{0}^{2}\right\rangle +\left\langle u_{1}^{2}\right\rangle\right] \frac{%
\partial \nu }{\partial r}{\rm d}r=0.  \label{kin}
\eea

\subsubsection{Geometrical quantities characterizing  galactic clusters}

In order to obtain some analytical estimations of the main geometrical and physical quantities characterizing the galactic clusters we introduce some approximations to the motion of both test particles in stable circular orbits around galaxies, as well as  to the motion of galaxies in galactic clusters. As a first approximation we assume that $\nu $ and $\lambda $ are slowly
varying functions of the radial coordinate $r$. 
Then  in Eq.~(\ref{ff1}) we can neglect all the quadratic terms as being negligibly small as compared to the first order terms. Secondly, astronomical observations show that the motion of the galaxies in clusters is non-relativistic. Hence the galactic velocities are much smaller
 than the velocity of the light, that is, $\langle u_{1}^{2}\rangle \approx \langle u_{2}^{2}\rangle \approx
\langle u_{3}^{2}\rangle \ll \langle u_{0}^{2}\rangle \approx 1$. Thus,
Eqs.~(\ref{ff1}) and (\ref{kin}) can be written as
\be  \label{fin1}
\frac{1}{2r^{2}}\frac{\partial }{\partial r}\left(r^{2} \frac{\partial \nu }{%
\partial r}\right) = 4\pi G\rho + 4\pi G\rho_{\phi}^{(eff)}\,,
\ee
and
\be
2K-\frac{1}{2}\int_{0}^{R}4\pi r^{3}\rho \frac{\partial \nu }{\partial r}%
{\rm d}r=0\,,  \label{cond1}
\ee
respectively, where
\begin{align}
K=\int_{0}^{R}2\pi \rho \left[ \left\langle u_{1}^{2}\right\rangle
+\left\langle u_{2}^{2}\right\rangle +\left\langle u_{3}^{2}\right\rangle %
\right] r^{2}{\rm d}r,
\end{align}
is the total kinetic energy of the galaxies in the cluster. We define the total baryonic mass  $M_B$ of the galactic cluster as
\begin{align}
 M_B=\int_{0}^{R}{\rm d}M(r)=\int_{0}^{R} 4\pi \rho r^{2}{\rm d}r
 \end{align}
 We assume that the main contribution to the baryonic mass $M_B$ is due to the presence of the intra-cluster
gas and of the stars. On the other hand we also include in $M_B$ the mass contribution of other particles, like, for example, massive neutrinos, which may
also give a significant contribution to $M_B$.

By multiplying Eq.~(\ref{fin1}) by $r^{2}$, and integrating from $0$ to $r$ we obtain
\begin{align}
GM_B(r)=\frac{1}{2}r^{2}\frac{\partial \nu }{\partial r}-GM_{\phi }^{(eff)}\left(
r\right),  \label{fin2}
\end{align}
where we have introduced the notation
\begin{align}  \label{darkmassb}
M_{\phi }^{(eff)}\left( r\right) =4\pi \int_{0}^{r}\rho _{\phi}^{(eff)}(r')r'^{2}
{\rm d}r'.
\end{align}

It is interesting to note at this point that in hybrid metric-Palatini gravity, the quantity $M_{\phi }^{(eff)}$ has essentially a geometric origin. Hence it is natural to call it as the \textit{geometric mass} of the galactic
cluster.  In  the following we introduce the gravitational potential energies of the cluster by means of the definitions
\begin{eqnarray}
\Omega _B&=&-\int_{0}^{R}\frac{GM_B(r)}{r}\,{\rm d}M_B(r)\,, \\
\Omega _{\phi }^{(eff)}&=&\int_{0}^{R}\frac{GM_{\phi }^{(eff)}(r)}{r}\,{\rm d}M_B(r)\,,
\end{eqnarray}
where $R$ is the cluster radius. By multiplying Eq.~(\ref{fin2}) with ${\rm d}M_B(r)$,  and by
integrating from $0$ to the cluster radius $R$, we obtain the important relation
\begin{align}
\Omega _B=\Omega _{\phi }^{(eff)}-\frac{1}{2}\int_{0}^{R}4\pi r^{3}\rho \frac{\partial
\nu }{\partial r}\,{\rm d}r\,.
\end{align}

\subsubsection{The generalized virial theorem in hybrid metric-Palatini gravity}

As a last step in our analysis, with the help of Eq.~(\ref{cond1}), we obtain  the generalization of
the virial theorem in hybrid metric-Palatini gravity, which can be formulated in the familiar form
\be
2K + \Omega = 0  \label{theor} \,.
\ee
In the above equation the total gravitational potential energy of the system, $\Omega$, defined as
\be
\Omega = \Omega _B- \Omega_{\phi }^{(eff)} \,, \label{theor2}
\ee
contains a supplementary term $\Omega_{\phi }^{(eff)} $, which has a purely geometric origin.

It is useful to represent the generalized virial theorem, given by Eq.~(\ref{theor}), in a more transparent physical  form, which can be obtained by introducing the radii $R_{V}$ and $R_{\phi }$, defined by
\be
R_{V}=M_B^{2}\Bigl/\int_{0}^{R}\frac{M_B(r)}{r}\,{\rm d}M_B(r),\Bigr.
\ee
and
\be
R_{\phi }^{(eff)}=\left[M_{\phi }^{(eff)}\right]^{2}\Bigl/\int_{0}^{R}\frac{M_{\phi }^{(eff)}(r)}{r}\,{\rm d}M_B(r),\Bigr.
\label{RU3}
\ee
respectively. The quantity $R_{\phi }$, having a geometric origin similarly to the geometric mass considered above,  may be called as the \textit{geometric radius} of the cluster of galaxies in the hybrid metric-Palatini gravity theory. Hence, finally we obtain the baryonic potential energy  $\Omega _B$ and the effective scalar field potential energy $\Omega _{\phi}^{(eff)}$ as given by
\be
\Omega _B =-\frac{GM_B^{2}}{R_{V}}\,, \quad
\Omega _{\phi}^{(eff)} =\frac{G\left[M_{\phi }^{(eff)}\right]^{2}}{R_{\phi }^{(eff)}}\,,
\ee
respectively. Another important observational quantity, the virial mass $M_{V}$ of the cluster of galaxies is defined as follows
\begin{align}\label{n1}
2K=\frac{GM_BM_{V}}{R_{V}}\,.
\end{align}

Eventually, the fundamental relation between the virial and the baryonic mass of the galaxy cluster can be obtained after substitution of Eq.~(\ref{n1}) into the virial theorem as
\be  \label{fin6}
\frac{M_{V}}{M_B}=1+\frac{\left[M_{\phi }^{(eff)}\right]^{2}R_{V}}{M_B^{2}R_{\phi }^{(eff)}}\,.
\ee
If $M_{V}/M_B>3$, a condition which is satisfied by the astrophysical parameters of most of the observed galactic
clusters, then from Eq.~(\ref{fin6}) we obtain the virial mass of galactic clusters in hybrid metric-Palatini gravity, which can be approximated as
\be
M_{V}\approx \frac{\left[M_{\phi}^{(eff)}\right]^2}{M_B}\frac{R_{V}}{R_{\phi }^{(eff)}}\,.  \label{virial}
\ee

The virial mass $M_V$ is determined observationally from the study of the velocity dispersion $\sigma _r^2$ of the stars and of the galaxies in the galactic clusters.  An important consequence  of the virial theorem in hybrid metric-Palatini gravity is that in a cluster with mass $M_{tot}$ most of the mass is in the form of the geometric mass $M_{\phi }^{(eff)}$.  Hence we can use the approximation  $M_{\phi}^{(eff)} \approx M_{tot}$ in order to study the cluster dynamics. A fundamental question related to the possibility of the observational testing of the hybrid metric-Palatini gravity is to find out what astrophysical processes can detect the presence of the geometric mass. Such an observational possibility  may be provided by gravitational lensing. Through the study of the lensing properties of the galactic clusters one can obtain direct evidence of the existence of the geometric mass, of its distribution properties, as well as of the gravitational effects associated to the presence of the scalar field. It is interesting to note that gravitational lensing can give us theoretical information even at cosmical scales extending far beyond of the virial radius of the matter distribution of the galactic cluster.

\subsubsection{On astrophysical tests of hybrid metric-Palatini gravity at the cluster level}\label{sec:d}

In concluding, in the framework of hybrid $f(X)$ gravity theory we have established the
the existence of a strict proportionality between the virial mass of the cluster and its baryonic mass, a relation which can also be tested observationally. One of the important, and observationally testable,  predictions of the hybrid metric-Palatini gravitational ``dark matter'' model is that the geometric masses associated to the clusters, as well as its gravitational  effects, extend beyond the virial radii of the clusters. Observationally, the virial mass $M_V$ is obtained from the study of the velocity dispersions of the stars in the cluster. Due to the observational uncertainties, this method  cannot give a reliable estimation of the numerical value of the total mass $M_B+M_{\phi }^{(eff)}$ in the cluster. However, a much more powerful method for the determination of the total mass distribution in clusters is the gravitational lensing of light, which may provide direct evidence for the gravitational effects at large distances from the cluster, and for the existence of the geometric mass. The presence of hybrid metric-Palatini modified gravity effects at large distances from the cluster, and especially the large extension of the geometric mass, may lead to significantly different lensing observational signatures, as compared to the standard relativistic/dark matter model case. The bending angle in the hybrid metric-Palatini gravity models could be larger
than the one predicted by the standard dark matter models. Therefore, the detailed observational study of the gravitational lensing could discriminate between the different theoretical models introduced  to explain the motion of galaxies (``particles") in the clusters of galaxies, and the standard dark matter models. We refer the reader to Ref. \cite{Capozziello:2012qt} for more details.

Finally, it is worth pointing out that hybrid gravity can be precisely tested also at smaller scales like those around the Galactic Centre. As reported in \cite{borka}, the observed peculiar orbit of S2 star, moving around the centre of our Galaxy,  is theoretically reconstructed if one adopts the gravitational potential coming from hybrid gravity. This result opens new perspectives in achieving precision tests for the theory.

\section{Conclusions}

In this work we have presented a hybrid metric-Palatini framework for theories of gravity, and have tested the new theories it entails using a number of theoretical consistency checks and observational constraints. From the field theory perspective, we found that the $f(X)$ class of theories, where $X=R+ \kappa^2T$, enjoys a similar special status amongst the more general hybrid metric-Palatini theories as the $f(R)$ theories within the narrower framework of purely metric gravity \cite{Woodard:2006nt}. This is so because when one excludes theories inhabited by ghost-like, superluminally propagating and otherwise pathological degrees of freedom, there is evidence, as shown in Section \ref{tts}, that the $f(X)$ family is singled out as the only viable form of an action one can construct using the metric (and thus the metric Levi-Civita connection) and an independent ``Palatini connection''. The underlying reason is that in the special case of $f(X)$ actions the higher derivatives in the gravity sector can be separated into a scalar mode, thus avoiding an Ostrogradskian instability. Indeed the $f(X)$ gravity represents a generic case within the one-parameter family of the Algebraic Scalar-Tensor theories (recall Eq. (\ref{eq:S_scalar2})), at one end of which lies the pure Palatini $f(\R)$ (wherein the scalar field is a function of the stress-energy trace $T$) and at the other end the pure metric $f(R)$ (where the field is a function of the metric curvature $R$). Furthermore, the propagating degrees of freedom have proven to be healthy also on curved backgrounds as confirmed also by our cosmological perturbation analysis in Section \ref{sec:pert}. Concerning the Cauchy problem, it was shown that in this class of theories the initial value problem can always be well-formulated and well-posed depending on the adopted matter sources. 

Having established the theoretical consistency and interest on the hybrid metric-Palatini $f(X)$ family of theories, we considered applications in which these theories provide gravitational alternatives to dark energy. As shown by our post-Newtonian analysis in Section \ref{weakfield}, the hybrid theories are promising in this respect as they can avoid the local gravity constraints but modify the cosmological dynamics at large scales. This is simply because as a scalar-tensor theory, the hybrid $f(X)$ gravity is characterised by an evolving Brans-Dicke coupling, which allows to introduce potentially large deviations from GR in the past (and future) as long as the coupling at the present epoch is strong enough to hide the field from the local gravity experiments. In contrast, in the metric $f(R)$ models the Brans-Dicke coupling is a finite constant and one needs to invoke some of the various ``screening mechanisms'' (workings of which remain to be studied in the hybrid theories) in order to reconcile the Solar system experiments with cosmology. 

Cosmological perturbations have been also analysed in these models up to the linear order \cite{Harko:2011nh,Capozziello:2012ny,Lima:2014aza}, and the results imply that the formation of large-scale structure in the aforementioned accelerating cosmologies is viable though exhibits subtle features that might be detectable in future experiments. In Section \ref{sec:pert}, we derived the full perturbations equations and extracting their Newtonian limit, describing the observable scales of the matter power spectrum, the growth of matter overdensities was shown to be modified by a time-dependent effective fifth force that is expected to modify the redshift evolution of the growth rate of perturbations. We also note that numerical studies of the perturbations imply that the difference of the gravitational potentials can exhibit oscillations at higher redshifts even when the background expansion and the full lensing potential are indistinguishable from the standard $\Lambda$CDM predictions \cite{Lima:2014aza}. Such features could potentially be observed in cross-correlations of the matter and lensing power spectra, but detailed calculations of the cosmic microwave background anisotropies and other perturbation observables remain to be carried out. This is especially worthwhile in view of the potential of the forthcoming Euclid mission to experimentally test different accelerating cosmological models by their impact to the large-scale  structure \cite{euclid}.

At an effective level, the $f(X)$ modifications involve both (the trace of) the matter stress energy and (the Ricci scalar of) the metric curvature, and from this point of view it appears appealing to speculate on the possible relevance of these theories to both the problems of dark energy and dark matter, in a unified theoretical framework and without distinguishing {\it a priori} matter and geometric sources. Various aspects of dark matter phenomenology from astronomical to galactic and extragalatic scales were discussed in section \ref{astro}. The generalised virial theoreom can acquire, in addition to the contribution from the baryonic masses, effective contributions of geometrical origin to the total gravitational potential energy, which may account for the well-known virial theorem mass discrepancy in clusters of galaxies. In the context of galactic rotation curves, the scalar-field modified relations between the various physical quantities such as tangential velocities of test particles around galaxies, Doppler frequency shifts and stellar dispersion velocities were derived. More recently, observational data of stellar motion near the Galactic centre was compared with simulations of the hybrid gravity theory, which turned out particularly suitable to model star dynamics. Yet, to promote the $f(X)$ theory into a convincing alternative to particle dark matter, one should produce also the cosmological successes of the $\Lambda$CDM model without the CDM component. 

Though wormhole solutions have already been found in these theories \cite{Capozziello:2012hr}, the nature of possible black hole solutions remains an outstanding open question. Though no-hair theorems for scalar-tensor theories exclude the simplest nontrivial solutions, their assumptions are restrictive. Another interesting though yet unstudied issue is the strong field regime of hybrid gravity and the constraints that one can put on the theories from astrophysical data such as measurements of binary pulsars. Finally, the possible relevance of the hybrid gravity framework to the early universe cosmology has not been explored in any detail. The Einstein frame formulation of the scalar-tensor description (\ref{einsteinframe}) could provide a useful starting point to investigate how the inflaton potential changes due to finite $\Omega_A$, in order to understand how the ''hybrid'' nature of spacetime modifies the standard predictions of e.g. the $\R^2$ model of inflation. \\

To conclude, whilst the physics of the metric and the Palatini versions of $f(R)$ gravity have been uncovered in exquisite detail in a great variety of different contexts \cite{fRgravity1,fRgravity2, fRgravity3, fRgravity4, fRgravity5, fRgravity6, fRgravity7, fRgravity8, fRgravity9,Olmo:2011uz}, those studies largely wait to be extended for the hybrid $f(X)$ version of the theory. We believe the results this far, as reported in this review,  provide compelling motivation for the further exploration of these particular theories.


\acknowledgments{Acknowledgments}

SC acknowledges the support of INFN ({\it iniziative specifiche} TEONGRAV and QGSKY). FSNL acknowledges financial support of the Fundac\~{a}o para a Ci\^{e}ncia e Tecnologia (Portugal) through an Investigador FCT Research contract, with reference IF/00859/2012, and the grants EXPL/FIS-AST/1608/2013, PEst-OE/FIS/UI2751/2014 and UID/FIS/04434/2013. GJO is supported by a Ramon y Cajal contract, the Spanish grant FIS2011-29813-C02-02, the Consolider Program CPANPHY-1205388, and the i-LINK0780 grant of the Spanish Research Council (CSIC).

\authorcontributions{Author Contributions}

All the authors have substantially contributed to the present work.

\conflictofinterests{Conflicts of Interest}

The authors declare no conflict of interest.

\bibliographystyle{mdpi}
\makeatletter
\renewcommand\@biblabel[1]{#1. }
\makeatother



\begin{thebibliography}{999} 

\bibitem{bah}
L. Iorio, 
``Editorial for the Special Issue 100 Years of Chronogeometrodynamics: The Status of the Einstein's Theory of Gravitation in Its Centennial Year,'' 
   Universe, vol. 1, no. 1, pp. 38-81, (2015)

\bibitem{expansion1}
S.~Perlmutter {\it et al.}  [Supernova Cosmology Project Collaboration],
  ``Measurements of Omega and Lambda from 42 high redshift supernovae,''
  Astrophys.\ J.\  {\bf 517}, 565 (1999)
  [astro-ph/9812133].

\bibitem{expansion2}
A.~G.~Riess {\it et al.}  [Supernova Search Team Collaboration],
  ``Observational evidence from supernovae for an accelerating universe and a cosmological constant,''
  Astron.\ J.\  {\bf 116}, 1009 (1998)
  [astro-ph/9805201].

\bibitem{expansion3}
A.~G.~Riess {\it et al.}  [Supernova Search Team Collaboration],
  ``Type Ia supernova discoveries at z > 1 from the Hubble Space Telescope: Evidence for past deceleration and constraints on dark energy evolution,''
  Astrophys.\ J.\  {\bf 607}, 665 (2004)
  [astro-ph/0402512].

\bibitem{expansion4}
A.~G.~Riess {\it et al.}  [Supernova Search Team Collaboration],
  ``The farthest known supernova: support for an accelerating universe and a glimpse of 
  the epoch of deceleration,''
  Astrophys.\ J.\  {\bf 560}, 49 (2001)
  [astro-ph/0104455].

\bibitem{expansion5}
 S.~Perlmutter, M.~S.~Turner and M.~J.~White,
  ``Constraining dark energy with SNe Ia and large scale structure,''
  Phys.\ Rev.\ Lett.\  {\bf 83}, 670 (1999)
  [astro-ph/9901052].


\bibitem{expansion6}
 C.~L.~Bennett {\it et al.}  [WMAP Collaboration],
  ``First year Wilkinson Microwave Anisotropy Probe (WMAP) observations: Preliminary 
  maps and basic results,''
  Astrophys.\ J.\ Suppl.\  {\bf 148}, 1 (2003)
  [astro-ph/0302207].

\bibitem{expansion7}
G.~Hinshaw {\it et al.}  [WMAP Collaboration],
  ``First year Wilkinson Microwave Anisotropy Probe (WMAP) observations: The Angular 
  power spectrum,''
  Astrophys.\ J.\ Suppl.\  {\bf 148}, 135 (2003)
  [astro-ph/0302217].


\bibitem{fRgravity1}
S.~Capozziello,
  ``Curvature quintessence,''
  Int.\ J.\ Mod.\ Phys.\ D {\bf 11}, 483 (2002)
  [gr-qc/0201033].

\bibitem{fRgravity2}
E.~J.~Copeland, M.~Sami and S.~Tsujikawa,
  ``Dynamics of dark energy,''
  Int.\ J.\ Mod.\ Phys.\ D {\bf 15}, 1753 (2006)
  [hep-th/0603057].
  
\bibitem{fRgravity3}
A.~De Felice and S.~Tsujikawa,
  ``$f(R)$ theories,''
  Living Rev.\ Rel.\  {\bf 13}, 3 (2010).
  [arXiv:1002.4928 [gr-qc]].
  
\bibitem{fRgravity4}
S.~M.~Carroll, V.~Duvvuri, M.~Trodden and M.~S.~Turner,
  ``Is cosmic speed - up due to new gravitational physics?,''
  Phys.\ Rev.\ D {\bf 70}, 043528 (2004)
  [astro-ph/0306438].
 
  
  \bibitem{fRgravity5}
 F.~S.~N.~Lobo,
  ``The dark side of gravity: Modified theories of gravity,''
  Dark Energy-Current Advances and Ideas, 173-204 (2009), Research Signpost, ISBN 978-81-308-0341-8
  [arXiv:0807.1640 [gr-qc]].
  
\bibitem{fRgravity6}
   S.~Capozziello and M.~De Laurentis,
  ``Extended Theories of Gravity,''
  Phys.\ Rept.\  {\bf 509}, 167 (2011)
  [arXiv:1108.6266 [gr-qc]].

\bibitem{fRgravity7} 
  S.~'i.~Nojiri and S.~D.~Odintsov,
  ``Introduction to modified gravity and gravitational alternative for dark energy,''
  Int.\ J.\ Geom.\ Meth.\ Mod.\ Phys.\  {\bf 4}, 115 (2007).
  
 \bibitem{fRgravity8} 
K.~Bamba, S.~Capozziello, S.~Nojiri and S.~D.~Odintsov,
  ``Dark energy cosmology: the equivalent description via different theoretical models and 
  cosmography tests,''
  Astrophys.\ Space Sci.\  {\bf 342}, 155 (2012)
  [arXiv:1205.3421 [gr-qc]].

 \bibitem{fRgravity9} 
S.~Nojiri and S.~D.~Odintsov,
  ``Unified cosmic history in modified gravity: from F(R) theory to Lorentz non-invariant 
  models,''
  Phys.\ Rept.\  {\bf 505}, 59 (2011)
  [arXiv:1011.0544 [gr-qc]].

\bibitem{Olmo:2011uz}
  G.~J.~Olmo,
  ``Palatini Approach to Modified Gravity: f(R) Theories and Beyond,''
  Int.\ J.\ Mod.\ Phys.\ D {\bf 20}, 413 (2011)
  [arXiv:1101.3864 [gr-qc]].
  
\bibitem{screen1} 
  A.~Joyce, B.~Jain, J.~Khoury and M.~Trodden,
  ``Beyond the Cosmological Standard Model,''
  Phys.\ Rept.\  {\bf 568}, 1 (2015)
  P.~Brax,
\bibitem{screen2}
  ``Screened modified gravity,''
  Acta Phys.\ Polon.\ B {\bf 43}, 2307 (2012)
  [arXiv:1211.5237 [hep-th]].
\bibitem{screen3}
 T.~S.~Koivisto, D.~F.~Mota and M.~Zumalacarregui,
  ``Screening Modifications of Gravity through Disformally Coupled Fields,''
  Phys.\ Rev.\ Lett.\  {\bf 109}, 241102 (2012)
  [arXiv:1205.3167 [astro-ph.CO]].
\bibitem{screen4} 
  P.~Brax, C.~van de Bruck, D.~F.~Mota, N.~J.~Nunes and H.~A.~Winther,
  ``Chameleons with Field Dependent Couplings,''
  Phys.\ Rev.\ D {\bf 82}, 083503 (2010)
  [arXiv:1006.2796 [astro-ph.CO]].
 
 
 \bibitem{screen5} 
  P.~Brax, A.~-C.~Davis, B.~Li and H.~A.~Winther,
  ``A Unified Description of Screened Modified Gravity,''
  Phys.\ Rev.\ D {\bf 86}, 044015 (2012)
  [arXiv:1203.4812 [astro-ph.CO]].

  
\bibitem{Koivisto:2005yc}
  T.~Koivisto and H.~Kurki-Suonio,
  ``Cosmological perturbations in the Palatini formulation of modified
  gravity,''
  Class.\ Quant.\ Grav.\  {\bf 23}, 2355 (2006)
  [arXiv:astro-ph/0509422].
  
\bibitem{Koivisto:2006ie} 
  T.~Koivisto,
  Phys.\ Rev.\ D {\bf 73}, 083517 (2006)
  [astro-ph/0602031].

\bibitem{Olmo:2008ye} 
  G.~J.~Olmo,
  ``Hydrogen atom in Palatini theories of gravity,''
  Phys.\ Rev.\ D {\bf 77}, 084021 (2008)
  [arXiv:0802.4038 [gr-qc]].


\bibitem{Olmo:2006zu} 
  G.~J.~Olmo,
  ``Violation of the Equivalence Principle in Modified Theories of Gravity,''
  Phys.\ Rev.\ Lett.\  {\bf 98}, 061101 (2007)
  [gr-qc/0612002].

\bibitem{Harko:2011nh}
  T.~Harko, T.~S.~Koivisto, F.~S.~N.~Lobo and G.~J.~Olmo,
  ``Metric-Palatini gravity unifying local constraints and late-time cosmic 
  acceleration,''
  Phys.\ Rev.\ D {\bf 85}, 084016 (2012)
  [arXiv:1110.1049 [gr-qc]].

\bibitem{IJMPD} 
S.~Capozziello, T.~Harko, F.~S.~N.~Lobo and G.~J.~Olmo,
  ``Hybrid modified gravity unifying local tests, galactic dynamics and late-time cosmic 
  acceleration,''
  Int.\ J.\ Mod.\ Phys.\ D {\bf 22}, 1342006 (2013)
  [arXiv:1305.3756 [gr-qc]].

\bibitem{Amendola:2010bk}
  L.~Amendola, K.~Enqvist, T.~Koivisto,
  ``Unifying Einstein and Palatini gravities,''
  Phys.\ Rev.\  {\bf D83}, 044016 (2011)
  [arXiv:1010.4776 [gr-qc]].

\bibitem{Koivisto:2013xza} 
  T.~S.~Koivisto, D.~F.~Mota and M.~Sandstad,
  ``Novel aspects of C-theories in Cosmology,''
  arXiv:1305.4754 [astro-ph.CO].


\bibitem{vari0}
  T.~S.~Koivisto,
  ``On new variational principles as alternatives to the Palatini method,''
  Phys.\ Rev.\ D {\bf 83}, 101501 (2011)
  [arXiv:1103.2743 [gr-qc]].  

\bibitem{varia} 
  A.~Baykal and O.~Delice,
  ``A Unified Approach to Variational Derivatives of Modified Gravitational Actions,''
  Class.\ Quant.\ Grav.\  {\bf 28}, 015014 (2011)
  [arXiv:1012.4246 [gr-qc]].

\bibitem{vari1}
  J.~Beltran Jimenez, A.~Golovnev, M.~Karciauskas and T.~S.~Koivisto,
  ``The Bimetric variational principle for General Relativity,''
  Phys.\ Rev.\ D {\bf 86}, 084024 (2012)
  [arXiv:1201.4018 [gr-qc]].

\bibitem{varib} 
  A.~Baykal and ?–.~Delice,
  ``Multi-Scalar-Tensor Equivalents for Modified Gravitational Actions,''
  Phys.\ Rev.\ D {\bf 88}, 084041 (2013)
  [arXiv:1308.6106 [gr-qc]].

\bibitem{vari2}
  M.~Sandstad, T.~S.~Koivisto and D.~F.~Mota,
  ``Non-locality of the C- and D-theories,''
  Class.\ Quant.\ Grav.\  {\bf 30}, 155005 (2013)
  [arXiv:1305.0695 [gr-qc]].

\bibitem{vari3}
 J.~Beltr?¡n Jim?©nez and T.~S.~Koivisto,
  ``Extended Gauss-Bonnet gravities in Weyl geometry,''
  Class.\ Quant.\ Grav.\  {\bf 31}, 135002 (2014)
  [arXiv:1402.1846 [gr-qc]].

\bibitem{vari4}
  A.~Golovnev, M.~Karciauskas and H.~J.~Nyrhinen,
  ``ADM Analysis of Gravity Models within the Framework of Bimetric Variational Formalism,''
  JCAP {\bf 1505}, no. 05, 021 (2015)
  [arXiv:1412.0637 [gr-qc]].

\bibitem{Koivisto:2005yk} 
  T.~Koivisto,
  ``Covariant conservation of energy momentum in modified gravities,''
  Class.\ Quant.\ Grav.\  {\bf 23}, 4289 (2006)
  [gr-qc/0505128].


\bibitem{Allemandi:2005qs} 
  G.~Allemandi, A.~Borowiec, M.~Francaviglia and S.~D.~Odintsov,
  Phys.\ Rev.\ D {\bf 72}, 063505 (2005)
  [gr-qc/0504057].

\bibitem{Olmo:2014sra} 
  G.~J.~Olmo and D.~Rubiera-Garcia,
  ``Brane-world and loop cosmology from a gravity-matter coupling perspective,''
  Phys.\ Lett.\ B {\bf 740}, 73 (2015)
  [arXiv:1405.7184 [hep-th]].


\bibitem{Harko:2010mv}
 T.~Harko and F.~S.~N.~Lobo,
  ``f(R,$L_{m}$) gravity,''
  Eur.\ Phys.\ J.\  C {\bf 70}, 373 (2010)
  [arXiv:1008.4193 [gr-qc]].
%

\bibitem{Bertolami:2007gv}
 O.~Bertolami, C.~G.~Boehmer, T.~Harko and F.~S.~N.~Lobo,
  ``Extra force in f(R) modified theories of gravity,''
  Phys.\ Rev.\ D {\bf 75}, 104016 (2007)
  [arXiv:0704.1733 [gr-qc]].

\bibitem{Bertolami:2007vu}
O.~Bertolami and J.~Paramos,
  ``Do f(R) theories matter?,''
  Phys.\ Rev.\ D {\bf 77}, 084018 (2008)
  [arXiv:0709.3988 [astro-ph]].


\bibitem{Bertolami:2008ab}
  O.~Bertolami, F.~S.~N.~Lobo and J.~Paramos,
  ``Non-minimum coupling of perfect fluids to curvature,''
  Phys.\ Rev.\  D {\bf 78}, 064036 (2008).
  [arXiv:0806.4434 [gr-qc]].

\bibitem{BPHL}
O.~Bertolami, J.~Paramos, T.~Harko and F.~S.~N.~Lobo,
  ``Non-minimal curvature-matter couplings in modified gravity,''
  arXiv:0811.2876 [gr-qc].

\bibitem{HKL}
 T.~Harko, T.~S.~Koivisto and F.~S.~N.~Lobo,
  ``Palatini formulation of modified gravity with a nonminimal curvature-matter 
  coupling,''
  Mod.\ Phys.\ Lett.\ A {\bf 26}, 1467 (2011)
  [arXiv:1007.4415 [gr-qc]].

\bibitem{Harko:2011kv}
  T.~Harko, F.~S.~N.~Lobo, S.~'i.~Nojiri and S.~D.~Odintsov,
  ``$f(R,T)$ gravity,''
  Phys.\ Rev.\ D {\bf 84}, 024020 (2011)
  [arXiv:1104.2669 [gr-qc]].

\bibitem{Haghani:2013oma} 
 Z.~Haghani, T.~Harko, F.~S.~N.~Lobo, H.~R.~Sepangi and S.~Shahidi,
   ``Further matters in space-time geometry: $f(R,T,R_{\mu\nu}T^{\mu\nu})$ gravity,''
  Phys.\ Rev.\ D {\bf 88}, 044023 (2013)
  [arXiv:1304.5957 [gr-qc]].
  

\bibitem{Odintsov:2013iba} 
S.~D.~Odintsov and D.~S\'{a}ez-G\'{o}mez,
  ``$f(R, T, R_{\mu\nu} T^{\mu\nu})$ gravity phenomenology and $\Lambda$CDM universe,''
  Phys.\ Lett.\ B {\bf 725}, 437 (2013)
  [arXiv:1304.5411 [gr-qc]].
  

\bibitem{Harko:2012ve}
  T.~Harko and F.~S.~N.~Lobo,
  ``Geodesic deviation, Raychaudhuri equation, and tidal forces in modified gravity 
  with an arbitrary curvature-matter coupling,''
  Phys.\ Rev.\ D {\bf 86}, 124034 (2012)
  [arXiv:1210.8044 [gr-qc]].

\bibitem{stabi1}
  N.~Tamanini and T.~S.~Koivisto,
  ``Consistency of nonminimally coupled $f(R)$ gravity,''
  Phys.\ Rev.\ D {\bf 88}, no. 6, 064052 (2013)
  [arXiv:1308.3401 [gr-qc]].

\bibitem{stabi2}
  I.~Ayuso, J.~Beltr?¡n Jim?©nez and ?.~de la Cruz-Dombriz,
  ``Consistency of universally nonminimally coupled $f(R,T,R_{μν}T^{μν})$ theories,''
  Phys.\ Rev.\ D {\bf 91}, no. 10, 104003 (2015)
  [arXiv:1411.1636 [hep-th]].


\bibitem{Capozziello:2012ny}
 S.~Capozziello, T.~Harko, T.~S.~Koivisto, F.~S.~N.~Lobo and G.~J.~Olmo,
  ``Cosmology of hybrid metric-Palatini f(X)-gravity,''
  JCAP {\bf 1304}, 011 (2013)
  [arXiv:1209.2895 [gr-qc]].


\bibitem{Olmo:2005zr} 
  G.~J.~Olmo,
  ``The Gravity Lagrangian according to Solar System experiments,''
  Phys.\ Rev.\ Lett.\  {\bf 95}, 261102 (2005)
  [gr-qc/0505101].

\bibitem{Olmo:2005hc} 
  G.~J.~Olmo,
  ``Post-Newtonian constraints on $f(R)$ cosmologies in metric and Palatini formalism,''
  Phys.\ Rev.\ D {\bf 72}, 083505 (2005)
  [gr-qc/0505135].

\bibitem{Koivisto:2009jn}
  T.~S.~Koivisto,
  ``Cosmology of modified (but second order) gravity,''
  AIP Conf.\ Proc.\  {\bf 1206}, 79 (2010)
  [arXiv:0910.4097 [gr-qc]].

  \bibitem{Cauchy-others1}
S.~Capozziello and S.~Vignolo,
  ``On the well formulation of the initial value problem of metric-affine f(R)-gravity,''
  Int.\ J.\ Geom.\ Meth.\ Mod.\ Phys.\  {\bf 6}, 985 (2009)
  [arXiv:0901.3136 [gr-qc]].
  
\bibitem{Cauchy-others1b}
S.~Capozziello and S.~Vignolo,
  ``The Cauchy problem for metric-affine f(R)-gravity in presence of perfect-fluid matter,''
  Class.\ Quant.\ Grav.\  {\bf 26}, 175013 (2009)
  [arXiv:0904.3686 [gr-qc]].

\bibitem{Cauchy-others1c}
S.~Capozziello and S.~Vignolo,
  ``The Cauchy problem for metric-affine f(R)-gravity in presence of a Klein-Gordon scalar field,''
  Int.\ J.\ Geom.\ Meth.\ Mod.\ Phys.\  {\bf 8}, 167 (2011)
  [arXiv:1003.4280 [gr-qc]].

\bibitem{Cauchy-others1d}
S.~Capozziello and S.~Vignolo,
  ``The Cauchy problem for f(R)-gravity: An Overview,''
  Int.\ J.\ Geom.\ Meth.\ Mod.\ Phys.\  {\bf 9}, 1250006 (2012)
  [arXiv:1103.2302 [gr-qc]].

\bibitem{Cauchy-others2}
G.~J.~Olmo and H.~Sanchis-Alepuz,
  ``Hamiltonian Formulation of Palatini f(R) theories a la Brans-Dicke,''
  Phys.\ Rev.\ D {\bf 83}, 104036 (2011)
  [arXiv:1101.3403 [gr-qc]].

\bibitem{yvonne4} 
Y.~Choquet--Bruhat,  2009, {\it General Relativity and the Einstein equations}, Oxford University Press Inc., New York.


\bibitem{Salgado} 
M.~Salgado,
  ``The Cauchy problem of scalar tensor theories of gravity,''
  Class.\ Quant.\ Grav.\  {\bf 23}, 4719 (2006)
  [gr-qc/0509001].


\bibitem{Capozziello:2013gza} 
  S.~Capozziello, T.~Harko, F.~S.~N.~Lobo, G.~J.~Olmo and S.~Vignolo,
  ``The Cauchy problem in hybrid metric-Palatini f(X)-gravity,''
  Int.\ J.\ Geom.\ Meth.\ Mod.\ Phys.\  {\bf 11}, no. 5, 1450042 (2014)
  [arXiv:1312.1320 [gr-qc]].

\bibitem{Wald}
R. M. Wald, {\em General Relativity} , (University of Chicago Press, Chicago, 1984).

\bibitem{Leray} J.~Leray, 1953,  {\it Hyperbolic differential equations}, Institute for Advanced Study Pub., Princeton.

\bibitem{yvonne2} 
Y.~Choquet--Bruhat, 1962, Cauchy problem, in Gravitation: an introduction to current research, (L.~Witten ed.), Wiley, New York.

\bibitem{yvonne} 
Y.~Four\'es--Bruhat, {\it Bull. de la S.M.F.}, {\bf 86}, 155, 1958.

\bibitem{Woodard:2006nt} 
  R.~P.~Woodard,
  ``Avoiding dark energy with 1/r modifications of gravity,''
  Lect.\ Notes Phys.\  {\bf 720}, 403 (2007)
  [astro-ph/0601672].

\bibitem{TT}
T.~S.~Koivisto and N.~Tamanini,
  ``Ghosts in pure and hybrid formalisms of gravity theories: A unified analysis,''
  Phys.\ Rev.\ D {\bf 87}, no. 10, 104030 (2013)
  [arXiv:1304.3607 [gr-qc]].

\bibitem{spin1}
  T.~Biswas, E.~Gerwick, T.~Koivisto and A.~Mazumdar,
  ``Towards singularity and ghost free theories of gravity,''
  Phys.\ Rev.\ Lett.\  {\bf 108}, 031101 (2012)
  [arXiv:1110.5249 [gr-qc]].

\bibitem{spin2}
  T.~Biswas, T.~Koivisto and A.~Mazumdar,
  ``Nonlocal theories of gravity: the flat space propagator,''
  arXiv:1302.0532 [gr-qc].

\bibitem{Flanagan:2003iw}
  E.~E.~Flanagan,
  ``Higher order gravity theories and scalar tensor theories,''
  Class.\ Quant.\ Grav.\  {\bf 21}, 417 (2003)
  [arXiv:gr-qc/0309015].

  \bibitem{Bo} 
N.~Tamanini and C.~G.~Boehmer,
  ``Generalized hybrid metric-Palatini gravity,''
  Phys.\ Rev.\ D {\bf 87}, no. 8, 084031 (2013)
  [arXiv:1302.2355 [gr-qc]].

\bibitem{Carloni:2015bua} 
  S.~Carloni, T.~Koivisto and F.~S.~N.~Lobo,
  ``A dynamical system analysis of hybrid metric-Palatini cosmologies,''
  arXiv:1507.04306 [gr-qc].

\bibitem{Boehmer:2013oxa}
  C.~G.~B\"{o}hmer, F.~S.~N.~Lobo and N.~Tamanini,
  ``Einstein static Universe in hybrid metric-Palatini gravity,''
  Phys.\ Rev.\ D {\bf 88}, no. 10, 104019 (2013)
  [Phys.\ Rev.\ D {\bf 88}, 104019 (2013)]
  [arXiv:1305.0025 [gr-qc]].

\bibitem{Lima:2014aza} 
  N.~A.~Lima,
  ``Dynamics of Linear Perturbations in the hybrid metric-Palatini gravity,''
  Phys.\ Rev.\ D {\bf 89}, 083527 (2014)
  [arXiv:1402.4458 [astro-ph.CO]].

\bibitem{Lima:2015nma} 
  N.~A.~Lima and V.~Smer-Barreto,
  ``Constraints on hybrid metric-Palatini models from background evolution,''
  arXiv:1501.05786 [astro-ph.CO].



\bibitem{Borowiec:2014wva} 
  A.~Borowiec, S.~Capozziello, M.~De Laurentis, F.~S.~N.~Lobo, A.~Paliathanasis, M.~Paolella and A.~Wojnar,
  ``Invariant solutions and Noether symmetries in Hybrid Gravity,''
  Phys.\ Rev.\ D {\bf 91}, 023517 (2015)
  [arXiv:1407.4313 [gr-qc]].

\bibitem{Ma:1995ey} 
  C.~P.~Ma and E.~Bertschinger,
  ``Cosmological perturbation theory in the synchronous and conformal Newtonian 
  gauges,''
  Astrophys.\ J.\  {\bf 455}, 7 (1995)
  [astro-ph/9506072].


\bibitem{Boisseau:2000pr} 
  B.~Boisseau, G.~Esposito-Farese, D.~Polarski and A.~A.~Starobinsky,
  ``Reconstruction of a scalar tensor theory of gravity in an accelerating universe,''
  Phys.\ Rev.\ Lett.\  {\bf 85}, 2236 (2000)
  [gr-qc/0001066].

\bibitem{coupled} 
  T.~S.~Koivisto, E.~N.~Saridakis and N.~Tamanini,
  ``Scalar-Fluid theories: cosmological perturbations and large-scale structure,''
  arXiv:1505.07556 [astro-ph.CO].

\bibitem{delaCruzDombriz:2008cp} 
  A.~de la Cruz-Dombriz, A.~Dobado and A.~L.~Maroto,
  Phys.\ Rev.\ D {\bf 77}, 123515 (2008)
  [arXiv:0802.2999 [astro-ph]].
  
  \bibitem{Abebe:2011ry} 
  A.~Abebe, M.~Abdelwahab, A.~de la Cruz-Dombriz and P.~K.~S.~Dunsby,
  Class.\ Quant.\ Grav.\  {\bf 29}, 135011 (2012)
  [arXiv:1110.1191 [gr-qc]].
  
  \bibitem{Llinares:2013qbh} 
  C.~Llinares and D.~Mota,
  Phys.\ Rev.\ Lett.\  {\bf 110}, no. 16, 161101 (2013)
  [arXiv:1302.1774 [astro-ph.CO]].
  
  \bibitem{Sawicki:2015zya} 
  I.~Sawicki and E.~Bellini,
  arXiv:1503.06831 [astro-ph.CO].
  
  


\bibitem{Cembranos:2008gj} 
  J.~A.~R.~Cembranos,
  Phys.\ Rev.\ Lett.\  {\bf 102}, 141301 (2009)
  [arXiv:0809.1653 [hep-ph]].
  
\bibitem{Arbuzova:2011fu} 
  E.~V.~Arbuzova, A.~D.~Dolgov and L.~Reverberi,
  JCAP {\bf 1202}, 049 (2012)
  [arXiv:1112.4995 [gr-qc]].

\bibitem{Koivisto:2011tp} 
  T.~S.~Koivisto,
  ``The post-Newtonian limit in C-theories of gravitation,''
  Phys.\ Rev.\ D {\bf 84}, 121502 (2011)
  [arXiv:1109.4585 [gr-qc]].
 
 \bibitem{bah2}
  L. Iorio, 
  ``Gravitational anomalies in the solar system?'' 
  International Journal of Modern Physics D, Volume 24, Issue 6, id. 1530015 (2015)

\bibitem{21cm} 
   J.~R.~Pritchard and A.~Loeb,
  ``21-cm cosmology,''
  Rept.\ Prog.\ Phys.\  {\bf 75}, 086901 (2012)
  [arXiv:1109.6012 [astro-ph.CO]].

\bibitem{dm}  
 V.~C.~Rubin, N.~Thonnard and W.~K.~Ford, Jr.,
 ``Rotational properties of 21 SC galaxies with a large range of luminosities and 
 radii, from NGC 4605 /R = 4kpc/ to UGC 2885 /R = 122 kpc/,''
  Astrophys.\ J.\  {\bf 238}, 471 (1980).

\bibitem{dm2}
M.~Persic, P.~Salucci and F.~Stel,
  ``The Universal rotation curve of spiral galaxies: 1. The Dark matter connection,''
  Mon.\ Not.\ Roy.\ Astron.\ Soc.\  {\bf 281}, 27 (1996)
  [astro-ph/9506004].

\bibitem{Nuc01}
 U.~Nucamendi, M.~Salgado and D.~Sudarsky,
  ``An Alternative approach to the galactic dark matter problem,''
  Phys.\ Rev.\ D {\bf 63}, 125016 (2001)
  [gr-qc/0011049].

\bibitem{Nuc01b}
T.~Harko,
  ``Galactic rotation curves in modified gravity with non-minimal coupling between 
  matter and geometry,''
  Phys.\ Rev.\ D {\bf 81}, 084050 (2010)
  [arXiv:1004.0576 [gr-qc]].

\bibitem{La03}
 K.~Lake,
  ``Galactic potentials,''
  Phys.\ Rev.\ Lett.\  {\bf 92}, 051101 (2004)
  [gr-qc/0302067].


\bibitem{Capozziello:2013yha} 
  S.~Capozziello, T.~Harko, T.~S.~Koivisto, F.~S.~N.~Lobo and G.~J.~Olmo,
  ``Galactic rotation curves in hybrid metric-Palatini gravity,''
  Astropart.\ Phys.\  {\bf 50-52}, 65 (2013)
  [arXiv:1307.0752 [gr-qc]].

\bibitem{Ar05}  
M.~Arnaud, ``X-ray observations of clusters of galaxies'',
in ``Background Microwave Radiation and Intracluster Cosmology'',
Proceedings of the International School of Physics ``Enrico Fermi'', edited
by F.~Melchiorri and Y.~Rephaeli, published by IOS Press, The Netherlands,
and Societ\`{a} Italiana di Fisica, Bologna, Italy, p.77, (2005).

\bibitem{ReBo02}  
 T.~H.~Reiprich and H.~Boehringer,
  ``The Mass function of an X-ray flux-limited sample of galaxy clusters,''
  Astrophys.\ J.\  {\bf 567}, 716 (2002)
  [astro-ph/0111285].



\bibitem{Sch01}
P.~Schuecker, H.~Boehringer, K.~Arzner and T.~H.~Reiprich,
  ``Cosmic mass functions from Gaussian stochastic diffusion processes,''
  Astron.\ Astrophys.\  {\bf 370}, 715 (2001)
  [astro-ph/0102439].

\bibitem{Sch01b}
M.~Baldi, V.~Pettorino, G.~Robbers and V.~Springel,
  ``Hydrodynamical N-body simulations of coupled dark energy cosmologies,''
  Mon.\ Not.\ Roy.\ Astron.\ Soc.\  {\bf 403}, 1684 (2010)
  [arXiv:0812.3901 [astro-ph]].

\bibitem{Sch01c}
 B.~Li and J.~D.~Barrow,
  ``N-Body Simulations for Coupled Scalar Field Cosmology,''
  Phys.\ Rev.\ D {\bf 83}, 024007 (2011)
  [arXiv:1005.4231 [astro-ph.CO]].


\bibitem{Capozziello:2012qt}
  S.~Capozziello, T.~Harko, T.~S.~Koivisto, F.~S.~N.~Lobo and G.~J.~Olmo,
  ``The virial theorem and the dark matter problem in hybrid metric-Palatini gravity,''
  JCAP {\bf 1307}, 024 (2013).
  arXiv:1212.5817 [physics.gen-ph].

\bibitem{HaCh07} 
 T.~Harko and K.~S.~Cheng,
  ``The Virial theorem and the dynamics of clusters of galaxies in the brane world 
  models,''
  Phys.\ Rev.\ D {\bf 76}, 044013 (2007)
  [arXiv:0707.1128 [gr-qc]].

\bibitem{Li66}  
R.~Maartens and S.~D.~Maharaj,
  ``Collision free gases in spatially homogeneous space-times,''
  J.\ Math.\ Phys.\  {\bf 26}, 2869 (1985).

\bibitem{Li66b}
 S.~Bildhauer,
  ``Transport Equations for Freely Propagating Photons in Curved Space-times: A Derivation 
  by Wigner Transformation,''
  Class.\ Quant.\ Grav.\  {\bf 6}, 1171 (1989).


\bibitem{Ja70}  
J.~C.~Jackson, Month.\ Not.\ R.\ Astr.\ Soc.\ \textbf{148}, 249 (1970).

\bibitem{borka}
 B. Borka, S. Capozziello, P. Jovanovic, and V. Borka Jovanovic, ``Probing hybrid modified gravity by stellar motion around Galactic Centre,''
 [arXiv:1504.07832 [gr-qc]] (2015).


\bibitem{euclid} 
  L.~Amendola {\it et al.} [Euclid Theory Working Group Collaboration],
  ``Cosmology and fundamental physics with the Euclid satellite,''
  Living Rev.\ Rel.\  {\bf 16}, 6 (2013)
  [arXiv:1206.1225 [astro-ph.CO]].
 


\bibitem{Capozziello:2012hr}
S.~Capozziello, T.~Harko, T.~S.~Koivisto, F.~S.~N.~Lobo and
G.~J.~Olmo,
  ``Wormholes supported by hybrid metric-Palatini gravity,''
  Phys.\ Rev.\ D {\bf 86}, 127504 (2012).
  
 


\end{thebibliography}


%


%

\end{document}